\pdfoutput=1
\newcommand*{\ATLASLATEXPATH}{latex/}
\documentclass[cernpreprint=true, coverpage=false, atlasdraft=false, texlive=2016, UKenglish]{\ATLASLATEXPATH atlasdoc}
\LEcontact{Pat Ward cpw1@hep.phy.cam.ac.uk}
\usepackage{\ATLASLATEXPATH atlaspackage}
\usepackage{\ATLASLATEXPATH atlasbiblatex}

\usepackage{multirow}

\usepackage{\ATLASLATEXPATH atlasphysics}
\graphicspath{{logos/}{figures/}}

\usepackage{PbPbPhotonJetPaper-defs}

\AtlasTitle{Measurement of photon--jet transverse momentum correlations in $5.02$~\TeV\ Pb+Pb and $pp$ collisions with ATLAS}

\AtlasVersion{2.4}
\AtlasAbstract{%
Jets created in association with a photon can be used as a calibrated
probe to study energy loss in the medium created in nuclear
collisions. Measurements of the transverse momentum balance between
isolated photons and inclusive jets are presented using integrated
luminosities of 0.49~nb$^{-1}$ of Pb+Pb collision data at
$\sqrt{s_\mathrm{NN}}=5.02$~\TeV and 25~pb$^{-1}$ of $pp$ collision
data at $\sqrt{s}=5.02$~\TeV recorded with the ATLAS detector at the
LHC. Photons with transverse momentum $63.1 < p_\mathrm{T}^{\gamma} <
200$~\GeV and $\left|\eta^{\gamma}\right| < 2.37$ are paired with all
jets in the event that have $p_\mathrm{T}^\mathrm{jet} > 31.6$~\GeV
and pseudorapidity $\left|\eta^\mathrm{jet}\right| < 2.8$. The
transverse momentum balance given by the jet-to-photon $p_\mathrm{T}$
ratio, $x_\mathrm{J\gamma}$, is measured for pairs with azimuthal
opening angle $\Delta\phi > 7\pi/8$. Distributions of the per-photon
jet yield as a function of $x_\mathrm{J\gamma}$,
$(1/N_\gamma)(\mathrm{d}N/\mathrm{d}x_\mathrm{J\gamma})$, are
corrected for detector effects via a two-dimensional unfolding
procedure and reported at the particle level. In $pp$ collisions, the
distributions are well described by Monte Carlo event generators. In
Pb+Pb collisions, the $x_\mathrm{J\gamma}$ distribution is modified
from that observed in $pp$ collisions with increasing centrality,
consistent with the picture of parton energy loss in the hot nuclear
medium. The data are compared with a suite of energy-loss models and
calculations.}

\author{The ATLAS Collaboration}

\AtlasRefCode{HION-2018-06}
\PreprintIdNumber{CERN-EP-2018-196}

 \AtlasJournal{Phys.\ Lett.\ B}
 \AtlasJournalRef{\PLB 789 (2019) 167}
 \AtlasDOI{10.1016/j.physletb.2018.12.023}

\AtlasCoverSupportingNote{Measurement of photon-jet correlations in Pb+Pb and $pp$ collisions at $\sqrt{s_\mathrm{NN}} = 5.02$~\TeV}{https://cds.cern.ch/record/2195847}
\AtlasCoverSupportingNote{Centrality determination in $\sqrt{s_\mathrm{NN}} = 5.02$~\TeV Pb+Pb collision data in 2015}{https://cds.cern.ch/record/2212936}

 \AtlasCoverCommentsDeadline{n/a}

\AtlasCoverAnalysisTeam{Dennis Perepelitsa, Peter Steinberg}

\AtlasCoverEdBoardMember{Martin Rybar~(chair)}
\AtlasCoverEdBoardMember{Claire Gwenlan}
\AtlasCoverEdBoardMember{Jose Ocariz}

\AtlasCoverEgroupEditors{atlas-ana-hion-2018-06-analysis-team@cern.ch}

\AtlasCoverEgroupEdBoard{atlas-ANA-HION-2018-06-editorial-board@cern.ch}

\hypersetup{pdftitle={ATLAS document},pdfauthor={The ATLAS Collaboration}}

\begin{document}

\maketitle

%\tableofcontents

\section{Introduction}

The energy loss of fast partons traversing the hot, deconfined medium
created in nucleus--nucleus collisions can be studied in a controlled
and systematic way through the analysis of jets produced in
association with a high transverse momentum (\pT) prompt
photon~\cite{Wang:1996yh,Wang:1996pe,Qin:2009bk,Renk:2009ur,Qin:2012gp,Wang:2013cia,Dai:2012am}.
At leading order in quantum chromodynamics, the photon and leading jet
are produced back-to-back in the azimuthal plane, with equal
transverse momenta. Measurements of prompt photon production in Au+Au
collisions at the Relativistic Heavy Ion Collider
(RHIC)~\cite{Afanasiev:2012dg} and Pb+Pb collisions at the Large
Hadron Collider (LHC)~\cite{HION-2012-02} have confirmed that, since
photons do not participate in the strong interaction, their production
rates are not modified by the medium~\cite{Arleo:2011gc}. Thus, photons
provide an estimate of the \pT and direction of the parton produced in
the initial hard-scattering before it has lost energy through
interactions with the medium. Measurements of jet production with
different requirements on the photon kinematics can therefore shed
light on how the absolute amount of parton energy loss depends on the
initial parton \pT.

Furthermore, photon--jet events offer a particularly useful way to
probe the distribution of energy lost by jets in individual events,
and are complementary to measurements such as the dijet \pT
balance~\cite{HION-2010-02,HION-2012-11,CMS-HIN-10-004}. Whereas those
measurements report the ratio of the transverse momenta of two
final-state jets, both of which may have lost energy, photon--jet
events provide an alternative system in which one high-\pT object is
certain to remain unaffected by the hot nuclear medium. Finally, jets
produced in association with a photon are more likely to originate
from quarks than those produced in dijet events at the same \pT. Thus,
when considered together with measurements of dijets or of inclusive
jet~\cite{HION-2011-02,HION-2013-06,CMS-HIN-11-013} and
hadron~\cite{HION-2011-03,CMS-HIN-15-015,Acharya:2018qsh} production
rates in Pb+Pb collisions, analysis of photon--jet events can help to
further constrain the flavour (i.e.\ quark versus gluon) dependence of
parton energy loss.

Studies of photon--hadron correlations, in which high-\pT hadrons are
used as a proxy for the jet, were first performed at
RHIC~\cite{Adare:2009vd,Adare:2012qi,STAR:2016jdz}, and measurements
using fully reconstructed jets have since begun at the
LHC~\cite{CMS-HIN-11-010,CMS-HIN-16-002}. In the LHC studies, the
distribution of the photon--jet azimuthal separation, $\Delta\phi$,
was found to be consistent with that in simulated photon--jet events
embedded into a heavy-ion background, and the jet-to-photon transverse
momentum ratio, $\xjg = \pTjet / \pTgamma$, was studied for inclusive
photon--jet pairs. The per-photon jet yield
$(1/N_\gamma)(\mathrm{d}N/\mathrm{d}\xjg)$ distribution was shifted to
significantly smaller values in Pb+Pb data.

In these previous measurements, the \xjg distributions in Pb+Pb events
were not corrected for detector resolution effects, which led to a
substantial broadening of the reported distributions in data. As a
result, qualitative comparisons with models or even with the analogous
distributions in proton--proton ($pp$) data could only be accomplished
by applying an additional smearing to the comparison distributions to
introduce detector effects. Recent measurements of dijet \pT
correlations~\cite{HION-2012-11} and inclusive jet fragmentation
functions at large longitudinal momentum fraction~\cite{HION-2015-05}
in Pb+Pb collisions used unfolding procedures to correct for
bin-migration effects and return the distributions to the particle
level, i.e.\ free from detector effects. 

%In these cases, fully correcting the data revealed non-trivial
%features in the distributions which would not otherwise be evident.

This Letter reports a study of photon--jet correlations in Pb+Pb
collisions at a nucleon--nucleon centre-of-mass energy
$\sqrt{s_\mathrm{NN}} = 5.02$~\TeV and $pp$ collisions at the same
centre-of-mass energy $\sqrt{s} = 5.02$~\TeV. The data were recorded
in 2015 with the ATLAS detector at the LHC and correspond to
integrated luminosities of $0.49$~nb$^{-1}$ and $25$~pb$^{-1}$,
respectively.  Events containing a prompt photon with $63.1 < \pTgamma
< 200$~\GeV and pseudorapidity $\left|\eta^\gamma\right| < 2.37$
(excluding the region $1.37 < \left|\eta^\gamma\right| < 1.52$) are
studied. The \pT balance of photon--jet pairs for jets with $\pTjet >
31.6$~\GeV and $\left|\eta^\mathrm{jet}\right| < 2.8$ which are
approximately back-to-back with the photon in the transverse plane,
$\Delta\phi > 7\pi/8$, is analysed through the per-photon yield of
jets as a function of \xjg, with all jets that meet this selection
requirement counted separately. In Monte Carlo simulations, the
fraction of photons paired with more than one jet rises from $1$\% to
$\approx15$\% over the reported photon \pT ranges. The particular
photon and jet \pT ranges used in the measurement are chosen to be
evenly spaced on logarithmic scales to facilitate the unfolding
procedure described below.

The yields are corrected via data-driven techniques for background
arising from combinatoric pairings of each photon with unrelated jets
in Pb+Pb events and from the contamination by neutral mesons in the
photon sample. The resulting \xjg distributions are corrected for the
effects of the experimental resolution on the photon and jet \pT via a
two-dimensional unfolding procedure similar to that used in
Ref.~\cite{HION-2012-11}. Due to higher-order effects, photon--jet
events do not generally have the back-to-back leading order topology
mentioned above. Thus the $pp$ data, which includes these effects,
provides the reference distributions against which to interpret the
results in Pb+Pb events. This Letter directly compares photon--jet
data in Pb+Pb and $pp$ events, and with Monte Carlo event generators
and analytic
calculations~\cite{Chen:2018fqu,Kang:2017xnc,KunnawalkamElayavalli:2016ttl,Casalderrey-Solana:2015vaa}.

\section{Experimental set-up}

The ATLAS experiment~\cite{PERF-2007-01} is a multipurpose particle
detector with a forward--backward symmetric cylindrical geometry and
nearly $4\pi$ coverage.\footnote{ATLAS uses a right-handed coordinate
  system with its origin at the nominal interaction point (IP) in the
  centre of the detector and the $z$-axis along the beam pipe. The
  $x$-axis points from the IP to the centre of the LHC ring, and the
  $y$-axis points upward.  Cylindrical coordinates $(r, \phi)$ are
  used in the transverse plane, $\phi$ being the azimuthal angle
  around the $z$-axis. The pseudorapidity is defined in terms of the
  polar angle $\theta$ as $\eta = -\ln \tan(\theta/2)$. Transverse
  momentum and transverse energy are defined as $\pT = p \sin\theta$
  and $\ET = E \sin\theta$, respectively. $\Delta{R}$ is defined as
  $\sqrt{ (\Delta\eta)^2 + (\Delta\phi)^2 }$.} This analysis relies on
the inner detector, the calorimeter and the data acquisition and
trigger system.

The inner detector comprises three major subsystems: the pixel
detector and the silicon microstrip tracker, which extend out to
$\left|\eta\right| = 2.5$, and the transition radiation tracker which
extends to $\left|\eta\right| = 2.0$. The inner detector covers the
full azimuth and is immersed in a 2~T axial magnetic field. The pixel
detector consists of four cylindrical layers in the barrel region and
three disks in each endcap region. The silicon microstrip tracker
comprises four cylindrical layers (nine disks) of silicon strip
detectors in the barrel (endcap) region.

The calorimeter is a large-acceptance, longitudinally-segmented
sampling detector covering $\left|\eta\right| < 4.9$ with
electromagnetic (EM) and hadronic sections. The EM calorimeter is a
lead/liquid-argon sampling calorimeter with an accordion-shaped
geometry. It is divided into a barrel region, covering
$\left|\eta\right| < 1.475$, and two endcap regions, covering $1.375 <
\left|\eta\right| < 3.2$. The EM calorimeter has three primary
sections, longitudinal in shower depth, called ``layers'', in the
barrel region and up to $\left|\eta\right| = 2.5$ in the end cap
regions. In the barrel and first part of the end cap
($\left|\eta\right| < 2.4$), with the exception of the regions $1.4 <
\left|\eta\right| < 1.5$, the first layer has a fine segmentation in
$\eta$ ($\Delta\eta = 0.003$--$0.006$) to allow the discrimination of
photons from the two-photon decays of $\pi^0$ and $\eta$ mesons. Over
most of the acceptance, the total material upstream of the EM
calorimeter ranges from $2.5$ to $6$ radiation lengths. In the
transition region between the barrel and endcap regions ($1.37 <
\left|\eta\right| < 1.52$), the amount of material rises to 11.5
radiation lengths, and thus this region is not used for the detection
of photons. The hadronic calorimeter is located outside the EM
calorimeter. It consists of a steel/scintillator-tile sampling
calorimeter covering $\left|\eta\right| < 1.7$ and a liquid-argon
calorimeter with copper absorber covering $1.5 < \left|\eta\right| <
3.2$.

The forward calorimeter (FCal) is a liquid-argon sampling calorimeter
located on either side of the interaction point. It covers $3.1 <
\left|\eta\right| < 4.9$ and each half is composed of one EM and two
hadronic sections, with copper and tungsten serving as the absorber
material, respectively. The FCal is used to characterise the centrality
of Pb+Pb collisions as described below. Finally, zero-degree
calorimeters (ZDC) are situated at large pseudorapidity,
$\left|\eta\right| > 8.3$, and are primarily sensitive to spectator
neutrons.

A two-level trigger system is used to select events, with a
first-level trigger implemented in hardware followed by a
software-based (high-level) trigger. Data for this measurement were
acquired using a high-level photon trigger~\cite{TRIG-2016-01}
covering the central region ($\left|\eta\right| < 2.5$). At the
first-level trigger stage, the transverse energy of EM showers is
computed within regions of $\Delta\phi\times\Delta\eta =
0.1\times0.1$, and those showers which satisfy an \ET threshold are
used to seed the high-level trigger stage. At this next stage,
reconstruction algorithms similar to those applied in the offline
analysis use the full detector granularity to form the final trigger
decision. The trigger was configured with an online photon-\pT
threshold of $30$~\GeV ($20$~\GeV) in the $pp$ (Pb+Pb) running period
and required the candidate photon to satisfy a set of loose criteria
for the electromagnetic shower shape~\cite{TRIG-2016-01}. For the
Pb+Pb data-taking, the high-level trigger included a procedure to
estimate and subtract the underlying event (UE) contribution to the
\ET\ measured in the calorimeter~\cite{HION-2012-02}, ensuring high
efficiency in high-activity Pb+Pb events.

In addition to the photon trigger, Pb+Pb data were recorded with
minimum-bias triggers; these events are used to characterise the
centrality of Pb+Pb collisions as described in
Section~\ref{sect:dataMC}. The minimum-bias triggers are based on the
presence of a minimum amount of approximately 50~\GeV\ of transverse
energy in all sections of the calorimeter system ($\left|\eta\right| <
3.2$) or, for events that do not meet this condition, on substantial
energy deposits in both ZDC modules and an inner-detector track
identified by the high-level trigger system.

\section{Data selection and Monte Carlo samples}
\label{sect:dataMC}

Photon--jet events in $pp$ and Pb+Pb collisions are initially selected
for analysis by the high-level triggers described above. The typical
number of interactions per bunch crossing in the $pp$ and Pb+Pb
data-taking were one and smaller than $10^{-4}$, respectively. Events
are required to satisfy detector and data-quality requirements, and to
contain a vertex reconstructed from tracks in the inner detector. An
additional requirement in Pb+Pb collisions, based on the correlation
of the signals in the ZDC and the FCal, is used to reject a small
number of recorded events consistent with two Pb+Pb interactions in
the same bunch crossing (pile-up)~\cite{HION-2016-04}. The pile-up
rate is largest in the most central events, where it is at most 0.1\%
and rejected with an efficiency greater than 98\%. No pile-up
rejection is applied in $pp$ collisions.

The centrality of Pb+Pb events is defined using the total transverse
energy measured in the FCal, evaluated at the electromagnetic scale
and denoted by \sumETPb. The same observable was used to characterise
2010 and 2011 Pb+Pb data at $\sqrt{s_\mathrm{NN}} =
2.76$~\TeV~\cite{HION-2011-05} and a similar procedure, based on Monte
Carlo Glauber modeling~\cite{Miller:2007ri}, is followed in 2015
data~\cite{HION-2016-07}. In this analysis, Pb+Pb events within five
centrality ranges are considered that represent $0$--$10$\% (largest
\sumETPb values and degree of nuclear overlap), $10$--$20$\%,
$20$--$30$\%, $30$--$50$\% and $50$--$80$\% (smallest \sumETPb values
and degree of nuclear overlap) of the population. The mean number of
participating nucleons in minimum-bias Pb+Pb collisions,
$N_\mathrm{part}$, ranges from $33.3\pm1.5$ in 50--80\% events to
$358.8\pm2.3$ in 0--10\% events.

%Figure~\ref{fig:cent} shows the \sumETPb distribution and \sumETPb
%ranges in minimum-bias Pb+Pb events which correspond to these
%centrality selections.

%\begin{figure}[!t]
%\centering
%\includegraphics[width=0.60\columnwidth]{Figure_CentRun2PbPb.pdf}
%\caption{\label{fig:cent} Measured \sumETPb distribution in
%  minimum-bias Pb+Pb collisions at $\sqrt{s_\mathrm{NN}} =
%  5.02$~\TeV. Alternating shaded and unshaded regions from the
%  large-\sumETPb end of the distribution denote the $0$--$10$\%,
%  $10$--$20$\%, $20$--$30$\%, $30$--$50$\%, and $50$--$80$\% centrality
%  ranges. }
%\end{figure}

%\begin{table}[h]
%\begin{center}
%\begin{tabular}{c|c|c|c}
%Centrality range  & $\ensuremath{\sum\!{E_\mathrm{T}}}\xspace$ range & $N_\mathrm{part}$ & $N_\mathrm{coll}$ \\ \hline 
%50--80\% & $0.0637$-$0.525$~TeV & $33.3 \pm 1.5$ & $48.3 \pm 3.5$ \\
%30--50\% & $0.525$--$1.37$~TeV & $109.2 \pm 2.5$ &  $265 \pm 16$ \\
%20--30\% & $1.37$--$2.05$~TeV  & $189.2 \pm 2.8$ & $6057 \pm 38$ \\
%10--20\% & $2.05$--$2.99$~TeV  & $264.1 \pm 2.9$ & $1003 \pm 66$  \\
%0--10\% & $> 2.99$~TeV         & $358.8 \pm 2.3$ & $1635 \pm 114$ 
%\end{tabular}
%\caption{Geometric parameters and their systematic uncertainty in
%  5.02~\TeV\ Pb+Pb data.
%\label{table:centrality} }
%\end{center}
%\end{table}

% MC

Monte Carlo simulations of $\sqrt{s} = 5.02$~\TeV $pp$ photon--jet
events are used to correct the data for bin migration and inefficiency
effects, and for comparison with distributions measured in $pp$
collision data. For all the samples described below, the generated
events were passed through a full \textsc{Geant}~4
simulation~\cite{Agostinelli:2002hh,SOFT-2010-01} of the ATLAS
detector under the same conditions present during data-taking and were
digitised and reconstructed in the same way as the data.

For the primary simulation samples, the \PYTHIA
8.186~\cite{Sjostrand:2007gs} generator was used with the NNPDF23LO
parton distribution function (PDF) set~\cite{Ball:2012cx}, and
generator parameters which were tuned to reproduce a set of
minimum-bias data (the ``A14''
tune)~\cite{ATL-PHYS-PUB-2014-021}. Both the direct and fragmentation
photon contributions are included in the simulation.  Six million $pp$
events were generated with a generator-level photon in the \pT range
$50$~\GeV to $280$~\GeV.  Additionally, a sample of 18 million events
were produced with the same generator, tune and PDF, and were overlaid
at the detector-hit level with minimum-bias Pb+Pb events recorded
during the 2015 run. The relative contribution of events in this
``data-overlay'' sample were reweighted on an event-by-event basis to
match the \sumET distribution observed in the photon--jet events in
Pb+Pb data selected for analysis. Thus the Pb+Pb simulation samples
contain underlying-event activity levels and kinematic distributions
of jets (used in the combinatoric photon--jet background estimation)
identical to those in data.

Additional samples of $0.3$ million $pp$ events and 6 million events
overlaid with Pb+Pb data were produced with the \SHERPA
2.1.1~\cite{Gleisberg:2008ta} generator using the CT10 PDF
set~\cite{Lai:2010vv}, as were $0.6$ million $pp$
\HERWIG~7~\cite{Bellm:2015jjp} events with the MMHT UE tune and PDF
set~\cite{Harland-Lang:2014zoa}. The \SHERPA samples were generated
with leading-order matrix elements for photon--jet final states with
up to three additional partons, which were merged with the \SHERPA
parton shower. The \HERWIG events were generated in a way that
includes the direct and fragmentation photon contributions. Both the
\SHERPA and \HERWIG samples were filtered for the presence of a photon
in the required kinematic region, and are used because they contain
different photon+multijet topological distributions and jet-flavour
compositions.

At generator level, photons are required to be isolated by requiring
the sum of the transverse energy carried by primary
particles\footnote{Primary particles are defined as those with a
  proper mean lifetime, $\tau$, exceeding $c\tau = 10$~mm. For the jet
  and isolation $E_\mathrm{T}$ measurements, muons and neutrinos are
  excluded from the definition.}  in a cone of size $\Delta{R} = 0.3$
around the photon, $E_\mathrm{T}^\mathrm{iso}$, to be smaller than
$3$~\GeV. In the analysis, the background subtraction, described
below, removes photons which pass the isolation cut in data but fail
this isolation requirement at the particle level. Jets are defined by
applying the anti-$k_{t}$
algorithm~\cite{Cacciari:2008gp,Cacciari:2011ma} with radius parameter
$R = 0.4$ to primary particles within $\left|\eta\right| < 4.9$. In
simulation, the jet flavour, i.e. whether it is quark- or
gluon-initiated, is defined as the flavour of the highest-\pT parton
that points to the generator-level jet~\cite{PERF-2011-03}.

\section{Event reconstruction}

\subsection{Photon reconstruction}

%Photons are reconstructed following a procedure used for previous
%measurements of isolated prompt photon production in Pb+Pb
%collisions~\cite{HION-2012-02}. It is similar to that used extensively
%in \pp collisions~\cite{STDM-2014-09,STDM-2016-08}, but is applied to
%the set of calorimeter cells after an event-by-event estimation and
%subtraction of the pile-up and UE contribution to the deposited energy
%in each cell~\cite{HION-2011-02}. Photon candidates are initially
%reconstructed from clusters of energy deposited in EM calorimeter
%cells. 
Photon candidates are reconstructed from clusters of energy deposited
in EM calorimeter cells, following a procedure used for previous
measurements of isolated prompt photon production in Pb+Pb
collisions~\cite{HION-2012-02}. The procedure is similar to that used
extensively in \pp collisions~\cite{STDM-2014-09,STDM-2016-08}, but is
applied to the calorimeter cells after an event-by-event estimation
and subtraction of the pile-up and UE contribution to the deposited
energy in each cell~\cite{HION-2011-02}. In Pb+Pb collisions, all
photon candidates are treated as if they were unconverted
photons. Photon identification is based primarily on shower shapes in
the calorimeter~\cite{ATL-PHYS-PUB-2016-014}, selecting those
candidates which are compatible with originating from a single photon
impacting the calorimeter. The measurement of the photon energy is
based on the energy collected in a small region of calorimeter cells
centred on the photon ($\Delta\eta\times\Delta\phi = 0.075\times
0.175$ in the barrel and $\Delta\eta\times\Delta\phi =
0.125\times0.125$ in the endcaps), and is corrected via a dedicated
calibration~\cite{ATL-PHYS-PUB-2016-015}, which accounts for upstream
losses and both lateral and longitudinal leakage. The sum of
transverse energy in calorimeter cells inside a cone size of
$\Delta{R} = 0.3$ centred on the photon candidate, excluding a small
central area of size $\Delta\eta\times\Delta\phi = 0.125\times0.175$,
is used to compute the isolation energy
$E_\mathrm{T}^\mathrm{iso}$. It is corrected for the expected leakage
of the photon energy into the isolation cone.

Reconstructed photon candidates are required to satisfy identification
and isolation criteria. The identification working point (called
``tight'') includes requirements on each of several shower-shape
variables~\cite{ATL-PHYS-PUB-2016-014}. These criteria reject
two-photon decays of neutral mesons using information in the finely
segmented first calorimeter layers, and reject hadrons which began
showering in the EM section using information from the hadronic
calorimeter. The isolation energy is required to be
$E_\mathrm{T}^\mathrm{iso} < 3$~\GeV\ in $pp$ collisions. In Pb+Pb
collisions, where UE fluctuations significantly broaden the
distribution of $E_\mathrm{T}^\mathrm{iso}$ values, this requirement
is set to approximately one standard deviation of the Gaussian-like
part of the distribution centred at zero, $E_\mathrm{T}^\mathrm{iso} <
8$~\GeV.

In simulation, prompt photons in $pp$ collisions have a total
reconstruction and selection efficiency greater than $90$\%. At low
$\pT \approx 60$~\GeV\ in the most central Pb+Pb collisions, this
efficiency is $\approx60$\%, rising with increasing \pT and in less
central collisions. In all events, the \pT scale, defined as the mean
ratio of measured photon \pT to the generator-level \pT, for photons
which satisfy these criteria is within $0.5$\% ($1$\%) of unity in the
barrel (endcap). The \pT resolution decreases from $3$\% to $2$\% over
the measured \pT range.

\subsection{Jet reconstruction}

Jets are reconstructed following the procedure previously used in
2.76~\TeV\ and 5.02~\TeV\ $pp$ and Pb+Pb
collisions~\cite{HION-2013-06,HION-2013-08,HION-2011-02}, which is
briefly summarised here. The anti-$k_{t}$
algorithm~\cite{Cacciari:2011ma} with $R=0.4$ is applied to energy
deposits in the calorimeter grouped into towers of size
$\Delta\eta\times\Delta\phi = 0.1\times0.1$. An iterative procedure,
based entirely on data, is used to obtain an event-by-event estimate
of the average $\eta$-dependent UE energy density, including that from
pile-up, while excluding from the estimate the contribution from jets
arising from a hard scattering. An updated estimate of the jet
four-momentum is obtained by subtracting the UE energy from the
constituent towers of the jet. This procedure is also applied to $pp$
data. The \pT values of the resulting jets are corrected for the
average calorimeter response using an $\eta$- and \pT-dependent
calibration derived from simulation. An additional correction, derived
from {\em in situ} studies of events with a jet recoiling against a
photon or $Z$ boson and from the differences between the heavy-ion
reconstruction algorithm and that normally used in the $13$~\TeV\ $pp$
data~\cite{PERF-2016-04}, is applied.  A final correction at the
analysis level is applied to correct for a deficiency in jet
calibration due to it being derived from an event sample with a
different jet flavour composition.

% , for which the UE subtraction removes the small energy density
% deposited from additional interactions in the same bunch crossing.

The distribution of reconstructed jet \pT values was studied in
simulation as a function of generator-level jet \pT. In $pp$ and Pb+Pb
collisions, the jet \pT scale is within $1$\% of unity. In $pp$
collisions, the jet \pT resolution decreases from $15$\% at $\pT
\approx 30$~\GeV to $10$\% at $\pT \approx 200$~\GeV. In Pb+Pb
collisions, the resolution at fixed jet \pT becomes worse in more
central collisions in a way consistent with the increasing magnitude
of UE fluctuations in the jet cone. In the most central events and at
the lowest jet-\pT values, the resolution reaches $50$\%. At high \pT,
the resolution asymptotically becomes centrality-independent and, at
$200$~\GeV, consistent with that in $pp$ collisions. More information
about the jet reconstruction and jet performance in this dataset may be found
in Ref.~\cite{HION-2017-10}.

\section{Data analysis}

\subsection{Photon purity and yield}
\label{ssec:photonpurity}

After applying the identification and isolation selection criteria in
$pp$ collisions, approximately $19\,500$, $7800$, $4100$ and $400$
photons are selected with $\pTgamma = 63.1$--$79.6$~\GeV,
$79.6$--$100$~\GeV, $100$--$158$~\GeV and $158$--$200$~\GeV,
respectively. In Pb+Pb collisions, the analogous yields are $15\,400$,
$6300$, $3500$ and $300$. These raw yields are determined as a
function of \pTgamma and are then corrected for background and for the
effects of \pT bin migration.

%Table~\ref{table:photoncounts} summarises the total number of
%identified, isolated photons initially selected in each \pTgamma
%range in the data.

First, the selected photon sample is corrected for the background
contribution, primarily from misidentified neutral hadrons. For each
\pTgamma and centrality range, the purity of prompt photons within
this range is estimated with a double-sideband
approach~\cite{HION-2012-02,STDM-2014-09,STDM-2016-08}, which is
summarised in the following.

In addition to the nominal selection, background-enhanced samples of
photon candidates are defined by selecting photons failing at least
one of four specific shower-shape requirements (referred to as the
``non-tight'' selection), or by requiring that they are not isolated
such that $E_\mathrm{T}^\mathrm{iso} > 5$~\GeV\ in $pp$ collisions or
$E_\mathrm{T}^\mathrm{iso} > 10$~\GeV\ in Pb+Pb collisions.  Regions
$A$ and $B$ are defined as those containing tight photons which are
isolated and non-isolated, respectively, with region $A$ corresponding
to the signal photon selection. Regions $C$ and $D$ contain non-tight
photons which are isolated and non-isolated, respectively. The number
of photon candidates in each region is generally a mixture of signal
and background photons, i.e. those arising from neutral mesons inside
jets. The $E_\mathrm{T}^\mathrm{iso}$ distribution for background
photons is expected to be the same for the tight and non-tight
selections such that the distribution of background photons
``factorises'' along isolation and identification axes. Separately,
the probability that a prompt photon is found in regions $B$, $C$ or
$D$ is determined from simulation. This information and the background
factorisation assumption is then applied to the data to determine the
purity of photons in region $A$, defined as the ratio of the number of
signal photons to all selected photons. The purity increases
systematically with \pTgamma over the measured \pT range. In $pp$
collisions, it rises from $\approx85$\% at $\pTgamma = 80$~\GeV\ to
more than $95$\% at $100$~\GeV, while in Pb+Pb collisions it is
typically $\approx75$--$90$\% over the same kinematic range.

The background-corrected prompt photon yields are then corrected for
the resolution of the \pTgamma measurement. This is performed by
comparing the yields, evaluated separately as a function of
reconstructed and generator-level \pT, in simulation. Given the
good \pT resolution, these differ by $2$\% at most, and this
small resulting correction is applied to the yields in data.

%\begin{table}[t]
%\begin{center}
%\begin{tabular}{c|c|c|c|c}
%\multirow{2}{*}{Data} & \multicolumn{4}{c}{Photon $p_\mathrm{T}$ range} \\ 
% & $63.1-79.6$~GeV & $79.6-100$~GeV & $100-158$~GeV & $158-200$~GeV \\  \hline 
%$pp$           & 19,495 & 7,754 & 4,099 & 368 \\ 
%50--80\% Pb+Pb & 738    & 322   & 154   & 17  \\ 
%30--50\% Pb+Pb & 2,708  & 1,022 & 571   & 47  \\ 
%20--30\% Pb+Pb & 2,685  & 1,031 & 569   & 72  \\ 
%10--20\% Pb+Pb & 3,779  & 1,560 & 875   & 89  \\ 
%0--10\% Pb+Pb  & 5,480  & 2,322 & 1,377 & 113 \\ 
%\end{tabular}
%\caption{Tight, isolated photon counts in $pp$ and Pb+Pb data for each
%  $p_\mathrm{T}^{\gamma}$ selection.
%\label{table:photoncounts} }
%\end{center}
%\end{table}

\subsection{Jet background subtraction}

The raw jet yields, measured as a function of $x_\mathrm{J\gamma}$,
are corrected for two background components using data-driven methods. The
corrections are performed separately for each \pTgamma
interval and separately in $pp$ collisions and Pb+Pb collisions of
different centrality ranges.

The first background arises from the combination of a high-\pT photon
with jets unrelated to the photon-producing hard scattering. These
include jets from separate hard parton--parton scatterings and UE
fluctuations reconstructed as jets. This background is negligible in
$pp$ collisions. Because of the inclusive jet selection in the
analysis, the combinatoric background is purely additive and can be
statistically subtracted after scaling to the total photon yield. The
combinatoric jet yields are determined in the data-overlay simulation,
by examining the yield of reconstructed jets separated from a
generator-level photon by $\Delta\phi > 7\pi/8$.  Reconstructed jets
that are not consistent with a generator-level jet, i.e. no
generator-level jet with $\pT > 20$~\GeV within $\Delta{R} < 0.4$, are
deemed to arise from the original Pb+Pb data event and are thus
labelled as ``combinatoric'' jets. The combinatoric jet yields are
subtracted from the measured $x_\mathrm{J\gamma}$ distributions in
data.

% Since simulated events are overlaid with Pb+Pb data events, they
% contain the same kinematic distribution of unrelated jets as do the
% photon--jet events in data, in addition to the same detector effects.

The second background is related to the estimated purity of the
selected photons. The \xjg yields for photon candidates in region $A$
contain an admixture of dijets, specifically jets correlated with
misidentified neutral mesons. Since these hadrons pass experimental
isolation requirements, they may be, for example, the leading fragment
inside a jet. The shape of this background in the \xjg distribution is
determined by repeating the analysis for photon candidates in region
$C$, since this region contains mostly neutral mesons that remain
isolated at the detector level. The resulting per-photon \xjg
distributions are scaled to match the number of background photons, as
determined above in Section~\ref{ssec:photonpurity}, and their yields
are statistically subtracted from the jet yields for photons in region
$A$.

\begin{figure*}[t]
\begin{center}
\includegraphics[width=0.32\textwidth]{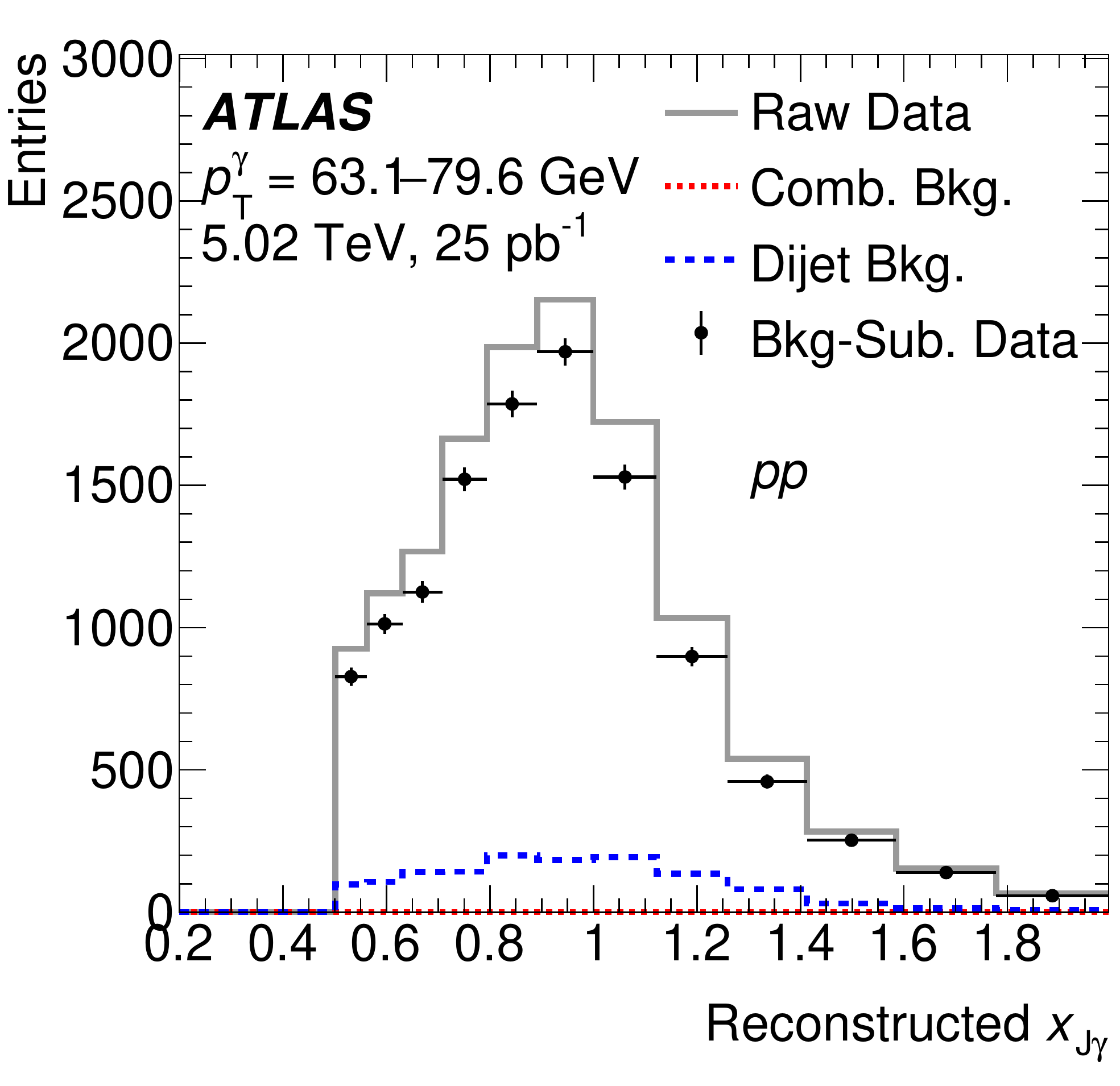}
\includegraphics[width=0.32\textwidth]{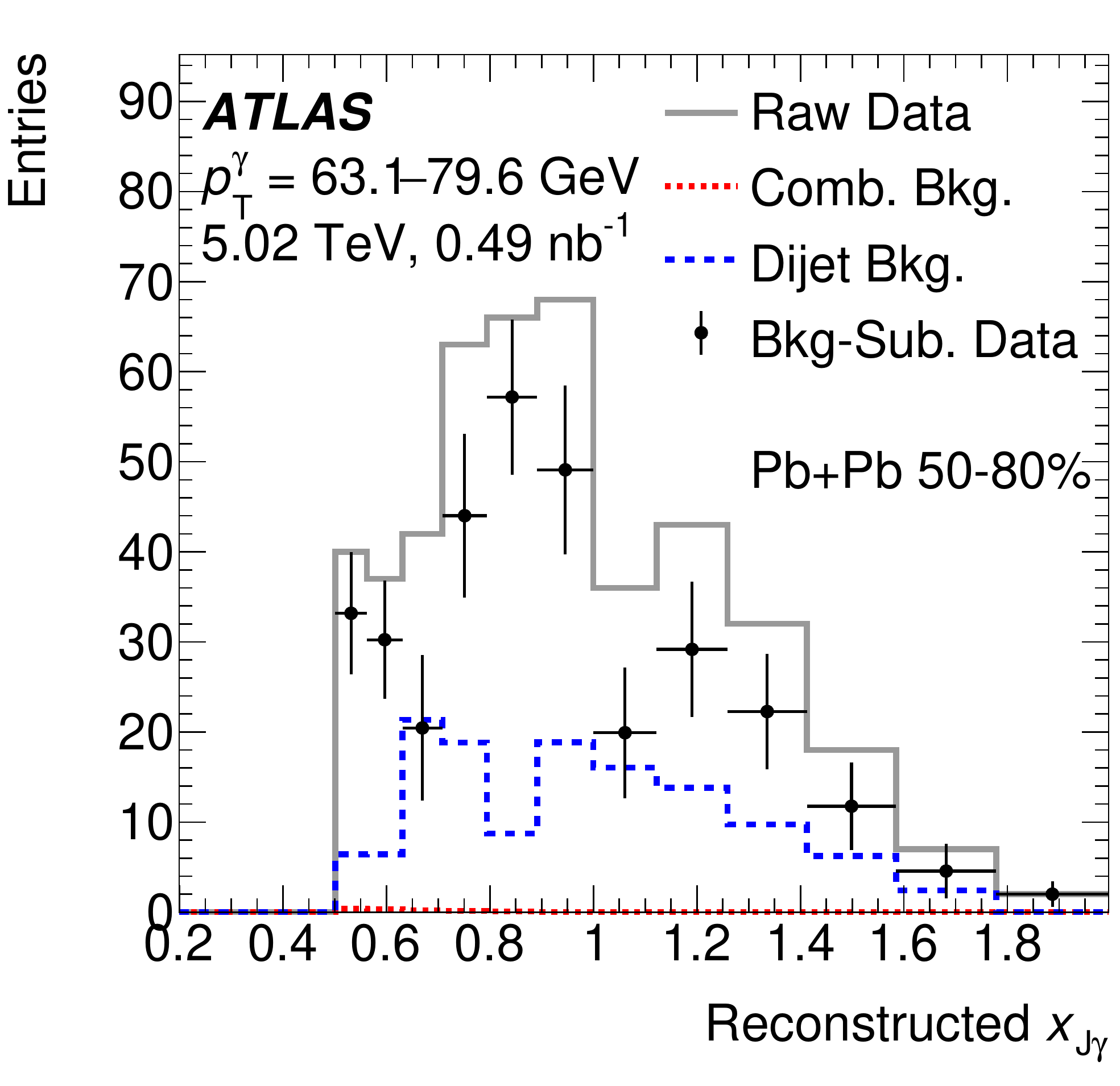}
\includegraphics[width=0.32\textwidth]{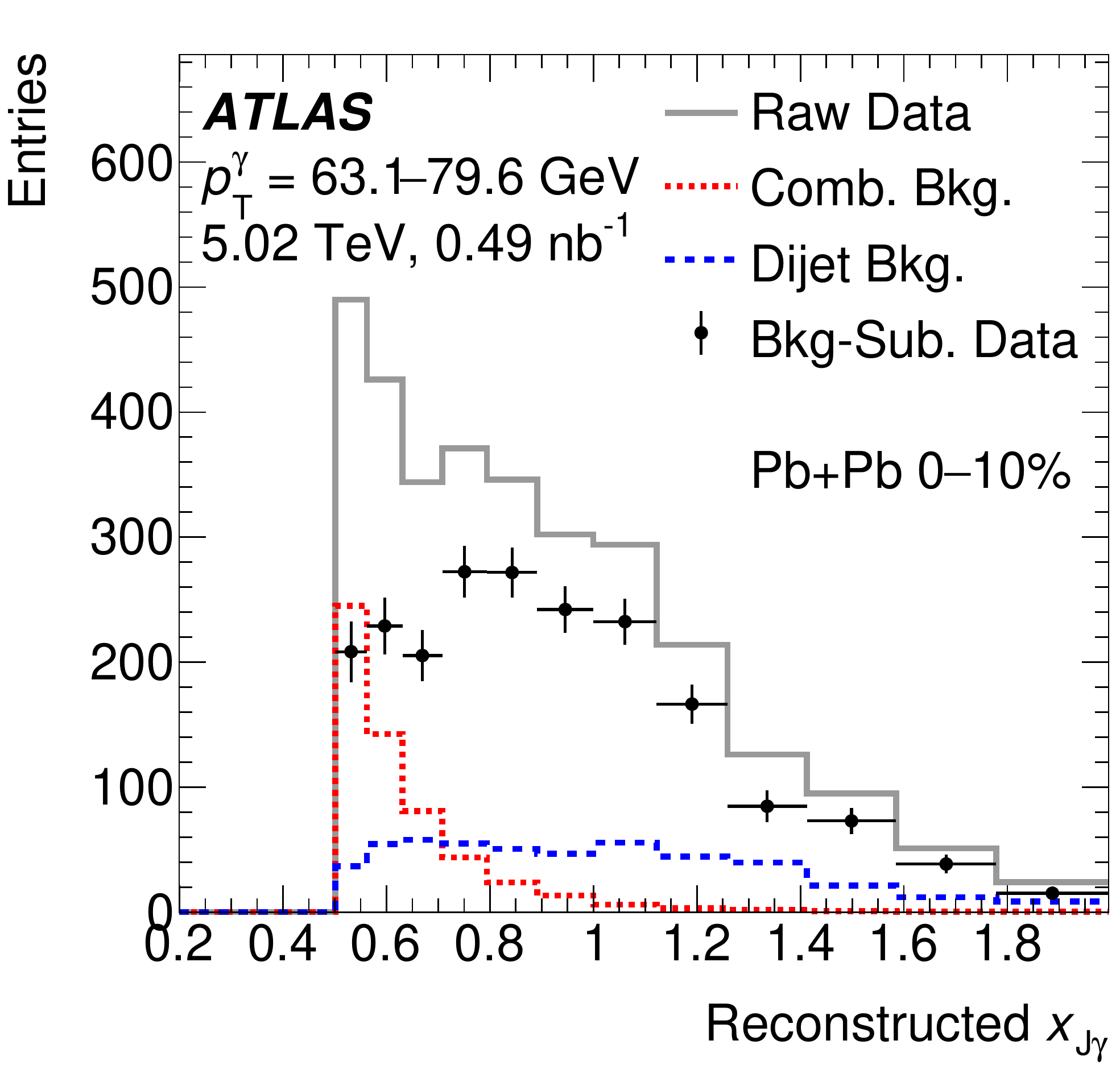}
\caption{Distributions of the photon--jet \pT-balance
  $x_\mathrm{J\gamma}$ for the photon transverse momentum interval
  $\pTgamma = 63.1$--$79.6$~\GeV for (left) $pp$, (centre) 50--80\%
  centrality and (right) 0--10\% centrality Pb+Pb events.  Solid grey,
  dotted red, and dashed blue histograms show the raw jet yields, the
  estimate of the combinatoric background (non-existent for $pp$
  events), and the dijet background, respectively. Black points show
  the background-subtracted data before unfolding, with the vertical
  bars representing the combined statistical uncertainty from the
  data and background subtraction procedure.
\label{fig:bkg}}
\end{center}
\end{figure*}

Figure~\ref{fig:bkg} shows the size of these backgrounds in the
lowest-\pTgamma interval, where they are the largest. The combinatoric
jet background for Pb+Pb collisions contributes primarily to kinematic
regions populated by $\pTjet < 50$~\GeV. It also depends strongly on
centrality, being largest in 0--10\% collisions but nearly negligible
already in 30--50\% collisions. The dijet background contributes to
a broad range of $\pTjet$ values including the region $\xjg > 1$,
since the \pT ratio of a jet to one of the hadrons in the balancing
jet can generally be above unity. This background has a similar shape
in all event types. However, since the photon purity is lower in Pb+Pb
events than in $pp$ events, this correction is larger in the former.

\subsection{Unfolding}

The background-subtracted \xjg yields are corrected for bin-migration
effects due to detector resolution via a Bayesian unfolding
procedure~\cite{DAgostini:1994fjx,Adye:2011gm}. To accomplish this,
the reconstructed yields are arranged in a two-dimensional $(\pTgamma,
\xjg)$ matrix with bin edges that are evenly spaced on logarithmic
scales (and with values matching those used in previous jet
measurements), and a two-dimensional unfolding is performed similar to
that for dijet \pT correlations in Ref.~\cite{HION-2012-11}. The
unfolding is performed in \xjg directly to preserve the fine
correlation between \pTjet and \pTgamma which would be washed out if
the unfolding were performed in $(\pTgamma, \pTjet)$. Although the
migration along the \pTgamma axis is small, it is necessary to include
it since the degree of bin migration in \xjg depends on the \pT of the
jets.

To fully account for the effects of bin migration across the analysis
selection, the axes of the matrix are extended over a larger range of
\pTgamma and \xjg than the fiducial region in which the results are
reported. A response matrix is determined by matching each pair of
$(\pTgamma, \xjg)$ values at the generator level to their
counterparts at the reconstruction level, separately for $pp$ events
and for each Pb+Pb centrality.

The Bayesian unfolding method requires a choice for the number of
iterations, $n_\mathrm{iter}$, and an assumption for the prior for the
initial particle-level distribution. The \PYTHIA simulation does not
include the effects of jet energy loss, and thus the underlying
particle-level distribution in data is expected to have a shape
different from the default prior in the simulation. An initial
unfolding using the default \PYTHIA prior is performed for each
centrality selection, and the ratios of the unfolded distributions to
the generator-level priors in \PYTHIA are fitted with a smooth
function in \xjg in each \pTgamma interval. This function is evaluated
to give a weight $w = w(\xjg, \pTgamma)$ that is used to reweight the
generator-level distribution in simulation and thus construct a
nominal prior. Alternative reweightings, used in evaluating the
sensitivity to the choice of prior, are determined by applying
$\sqrt{w}$ (the geometric mean of the nominal reweighting and no
reweighting) and $w^{3/2}$ to the sample. The reconstruction-level \xjg
distributions in simulation after each of these reweightings were
examined to ensure that they span a reasonable range of values
compared to that observed at the reconstruction level in data.

% An example of the sensitivity of the unfolded data to these priors
% is shown in Fig.~\ref{fig:unfsens} for the lowest \pTgamma bin in
% the most central events, where the reweighting is largest. While a
% large number of iterations should decrease the sensitivity to the
% choice of prior, this may amplify statistical uncertainties in the
% measurement more than is necessary.

Before applying the unfolding procedure to data, it was tested on
simulation. After the nominal reweighting, the Monte Carlo samples
were split into two statistically independent subsamples. One
subsample was used to populate the response matrix, which was then
used to unfold the reconstruction-level distribution in the other
subsample. The unfolded result was compared with the original
generator-level distribution in the latter sample, which were found to
be recovered within the limits of the statistical precision of the
samples.

The values of $n_\mathrm{iter}$ used for the nominal results are
chosen following the same procedure as in
Ref.~\cite{HION-2012-11}. For each centrality selection, the unfolded
distributions are examined as a function of $n_\mathrm{iter}$. For
each value of $n_\mathrm{iter}$, a total uncertainty is formed by
adding two components in quadrature: (1) the statistical uncertainty
of the unfolded data, which grows slowly with $n_\mathrm{iter}$, and
(2) the sum of square differences between the results and those
obtained with an alternative prior, which decreases quickly with
$n_\mathrm{iter}$. The final values of $n_\mathrm{iter}$ are chosen to
minimise the total uncertainty, and are between two and four.

The unfolded \xjg results are corrected for the jet reconstruction
efficiency, evaluated in simulation as the \pTgamma-dependent
probability that a generated jet at the given \xjg is successfully
reconstructed within the total $(\pTgamma,\xjg)$ range used in the
unfolding. This efficiency is typically $>99$\% for all events in the
kinematic regions populated by jets with $\pT > 50$~\GeV. In $pp$
collisions, this efficiency falls to $\approx96$\% in the lowest-\xjg
region for each \pTgamma interval. In Pb+Pb collisions, the efficiency
at fixed \xjg decreases monotonically in increasingly central events,
reaching a minimum of $\approx75$\% in the lowest-\xjg region in
0--10\% centrality events.

\section{Systematic uncertainties}

The primary sources of systematic uncertainty can be grouped into
three major categories: the measurement of \pTjet; the selection of
the photon and measurement of \pTgamma; the modelling and subtraction
of the combinatoric background; and the unfolding procedure.  For each
variation described below, the entire analysis is repeated including
the background correction steps and unfolding. The differences between
the resulting $x_\mathrm{J\gamma}$ values and the nominal ones are
taken as an estimate of the uncertainty from each source.

A standard set of uncertainties in the jet \pT scale and resolution,
following the strategy described in Ref.~\cite{ATLAS-CONF-2015-016}
and commonly used for measurements in 2015 Pb+Pb and $pp$
data~\cite{HION-2017-10,HION-2017-04}, are used in this analysis. The
impact of the uncertainties is evaluated by modifying the response
matrix according to the given variations in the reconstructed
jet \pT. These include uncertainties in the \pT scale derived from {\em
  in situ} studies of the calorimeter
response~\cite{PERF-2011-03,ATL-PHYS-PUB-2015-015}, an uncertainty in
the resolution derived using data-driven
techniques~\cite{ATLAS-CONF-2015-017}, and uncertainties in both which
result from a small relative energy-scale difference between the
heavy-ion jet reconstruction procedure and that used in $\sqrt{s} =
13$~\TeV $pp$ collisions~\cite{PERF-2016-04}. All of the above
uncertainties apply equally to jets in $pp$ and Pb+Pb events. A
separate, centrality-dependent uncertainty is included in 0--60\%
Pb+Pb collisions. This uncertainty accounts for a possible
modification of the jet response after energy loss and is evaluated
through {\em in situ} comparisons of the charged-particle track-jet
and calorimeter-jet \pT values in data and simulation. More details
are provided in Refs.~\cite{HION-2017-10,ATLAS-CONF-2015-016}. No
additional uncertainty is included for 60--80\% centrality events.

Uncertainties in the photon purity estimate are determined by varying
the non-tight identification and isolation criteria used to select
hadron background candidates and by considering a possible
non-factorisation of the hadron background along the axes used in the
double-sideband procedure. The sensitivity to the modelling of photon
shower shapes in simulation is evaluated by removing the data-driven
corrections to these quantities~\cite{ATL-PHYS-PUB-2016-014}. Finally,
the photon \pT scale and resolution uncertainties are described in
detail in Ref.~\cite{ATL-PHYS-PUB-2016-015}, and their impact is
evaluated by applying them as variations to the response matrices used
in unfolding.

Modelling- or unfolding-related systematic uncertainties arise from
several sources. The estimate of the combinatoric photon--jet rate in
the data-overlay simulation is sensitive to the requirement on the
minimum \pT of a generator-level jet in the classification of a given
reconstructed jet as a combinatoric jet, as opposed to a
photon-correlated jet. To provide one estimate of the sensitivity to
this threshold, it is varied in the range $20\pm10$~\GeV. To assess
the sensitivity to the choice of prior, the unfolding is repeated
using the alternative priors which are systematically closer to and
farther from the original \PYTHIA prior. The sensitivity to
statistical limitations of the simulation samples is determined
through pseudo-experiments, resampling entries in the response
matrices according to their uncertainty. Finally, the analysis is
repeated using the \SHERPA simulation to perform the corrections and
unfolding, since this generator provides a different description of
photon--jet production topologies.

\begin{figure*}[t]
\begin{center}
\includegraphics[width=0.49\textwidth]{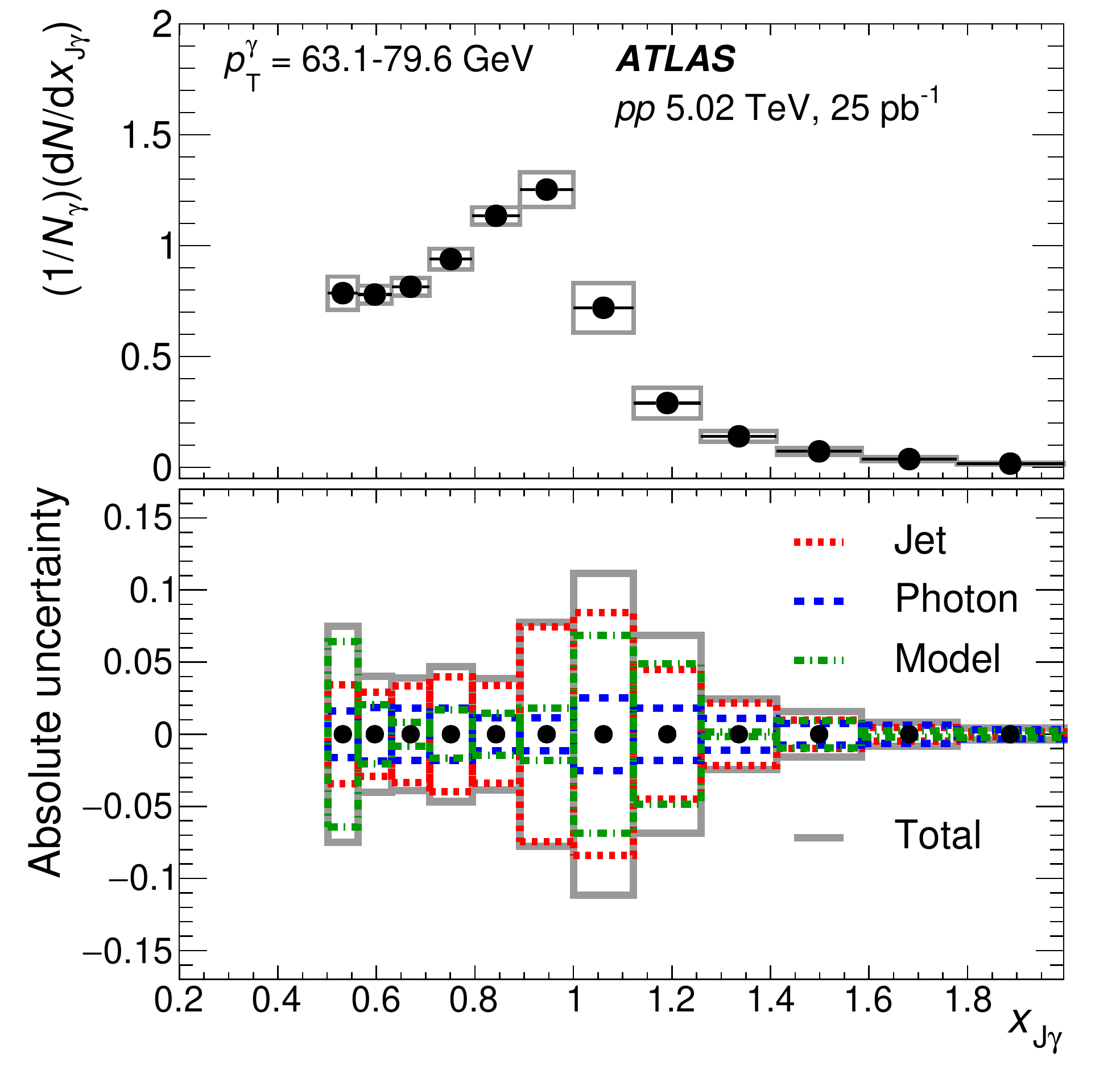}
\includegraphics[width=0.49\textwidth]{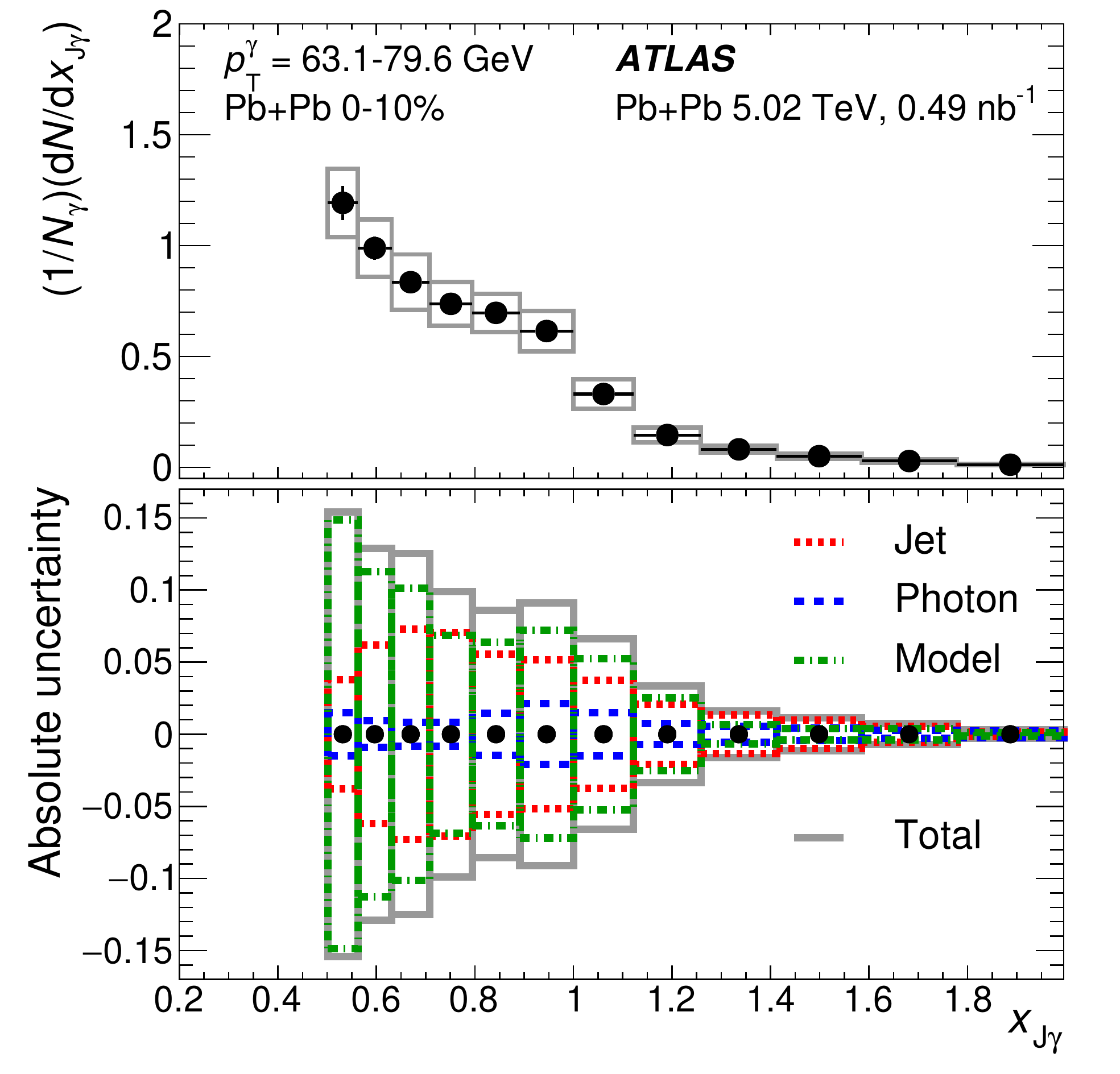}
\caption{Unfolded distributions and summary of systematic
  uncertainties in the per-photon jet-yield measurement for $\pTgamma
  = 63.1$--$79.6$~\GeV in (left) $pp$ events and (right) 0--10\%
  centrality Pb+Pb events. Top panels show the photon--jet \pT-balance
  \xjg\ distributions and total uncertainties, while the bottom panels
  show the absolute uncertainties from jet-related, photon-related,
  and modelling or unfolding sources, as well as the total
  uncertainty.
\label{fig:syst}}
\end{center}
\end{figure*}

Figure~\ref{fig:syst} summarises the systematic uncertainties in each
category, as well as the total uncertainty, for the lowest-\pTgamma
interval in $pp$ and 0--10\% Pb+Pb events. The jet-related
uncertainties are generally the dominant ones, except in more central
events and lower-\pTgamma intervals, where the unfolding and modelling
uncertainties become co-dominant.

As an additional check on the features in the unfolded \xjg
distributions observed in data, the analysis was repeated with two
modifications which change the signal photon--jet definition. First,
the photon--jet $\Delta\phi$ requirement was changed from $> 7\pi/8$
to $> 3\pi/4$. With this alteration, the correlated jet yield changes
only by a small amount, while the combinatoric background, which is
constant in $\Delta\phi$, doubles. Second, the analysis was repeated,
but selecting only the leading (highest-\pT) jet in the event if it
fell within the $\Delta\phi$ window. In this case, the combinatoric
background contribution is no longer purely additive and the
inefficiency when a higher-\pT uncorrelated jet is selected instead of
the photon-correlated jet must be accounted for, similar to
Ref.~\cite{HION-2012-11}. In both cases, the distributions in Pb+Pb
exhibit a qualitatively similar modification pattern compared to the
main results as a function of \xjg.

\section{Results}

\begin{figure*}[p]
\begin{center}
\includegraphics[width=0.49\textwidth]{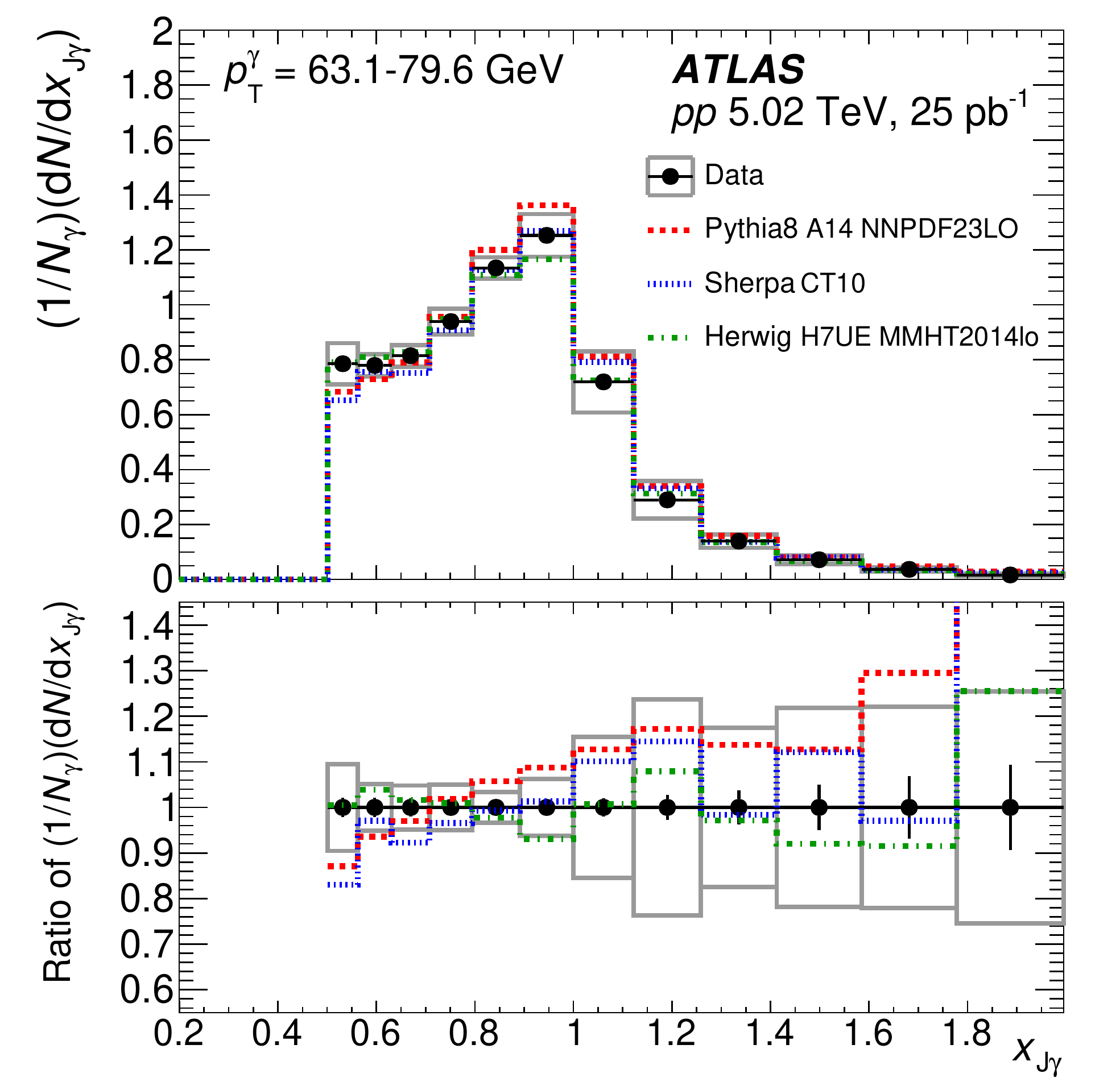}
\includegraphics[width=0.49\textwidth]{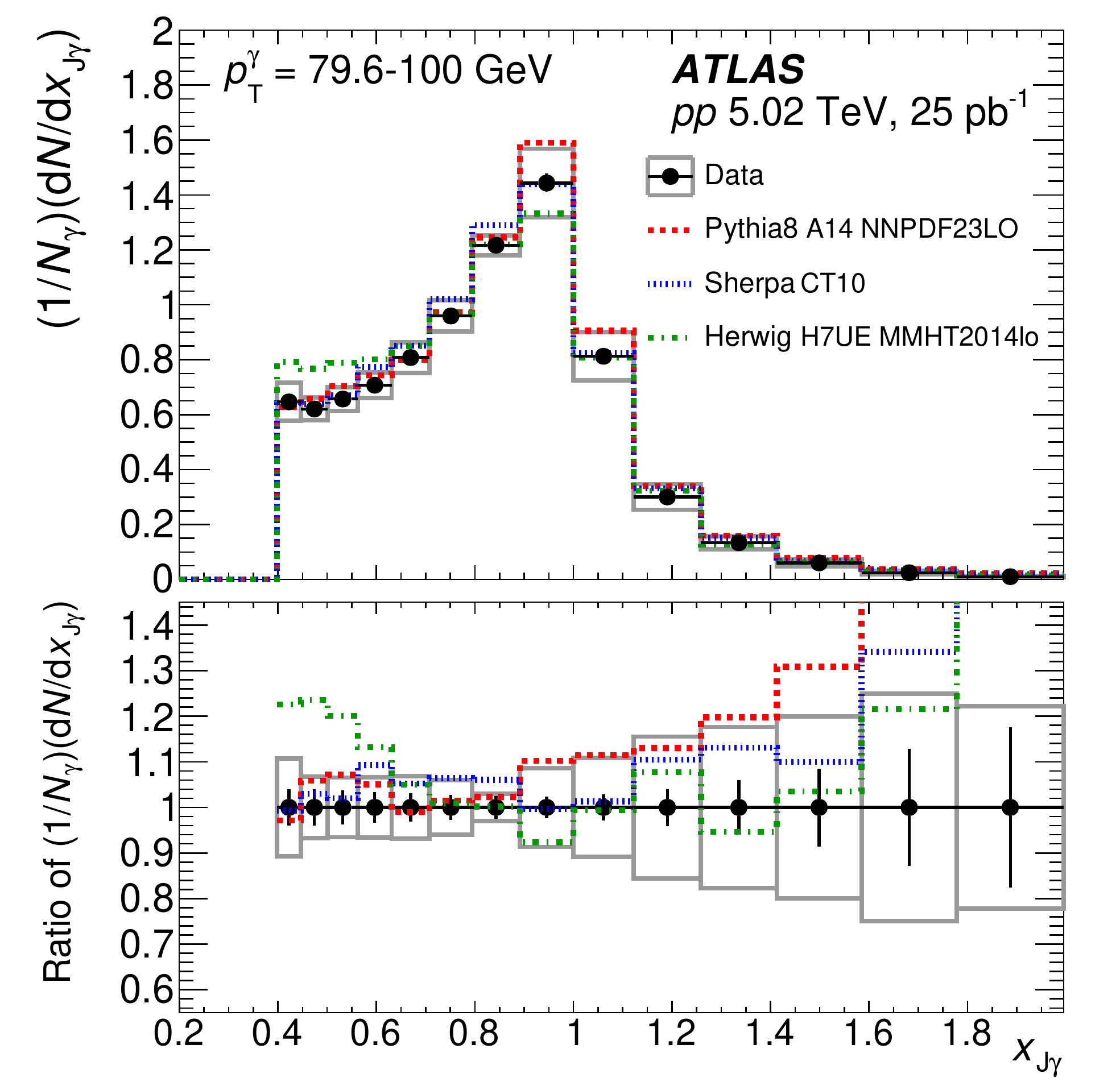}
\includegraphics[width=0.49\textwidth]{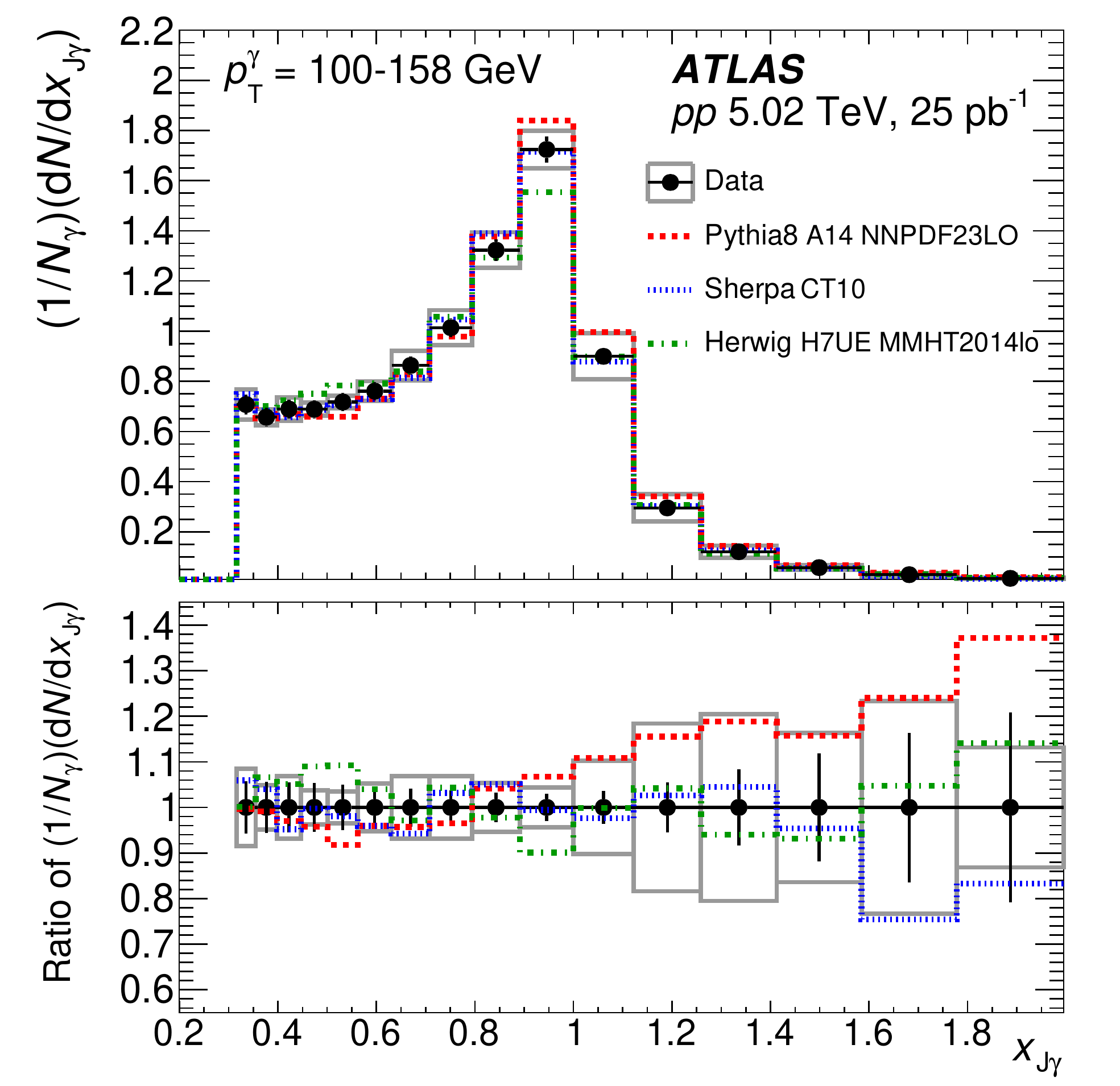}
\includegraphics[width=0.49\textwidth]{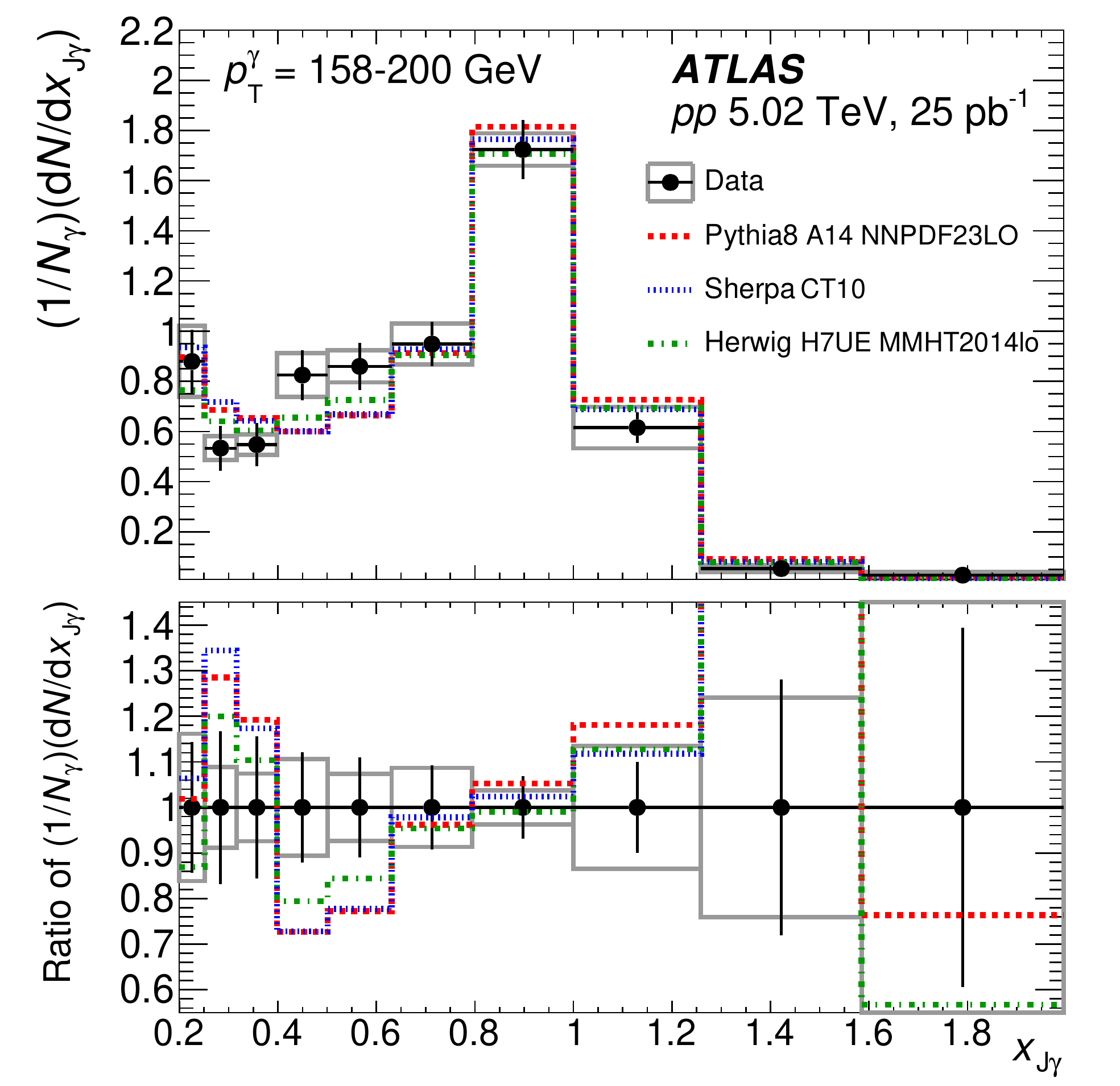}
\caption{Photon--jet \pT-balance distributions
  $(1/N_\gamma)(\mathrm{d}N/\mathrm{d}x_\mathrm{J\gamma})$ in $pp$
  collisions, each panel showing a different photon-\pT
  interval. The unfolded results are compared with the particle-level
  distributions from three Monte Carlo event generators. Bottom panels show the
  ratios of the generators to the $pp$ data. Total systematic
  uncertainties are shown as boxes, while statistical uncertainties
  are shown as vertical bars.
\label{fig:pp}}
\end{center}
\end{figure*}

The unfolded $(1/N_\gamma)(\mathrm{d}N/\mathrm{d}x_\mathrm{J\gamma})$
distributions in $pp$ collisions are shown for each \pTgamma interval
in Figure~\ref{fig:pp}. The distributions are reported for all \xjg
bins where the jet minimum \pT requirement is fully efficient. Also
shown are the corresponding generator-level distributions from the
\PYTHIA, \SHERPA and \HERWIG samples. Each generator describes the
data fairly well, with \HERWIG generally overpredicting the yield at
large-\xjg and \SHERPA showing the best agreement over the full \xjg
range.

%% include Pb+Pb results

\begin{figure*}[p]
\begin{center}
\includegraphics[width=0.99\textwidth]{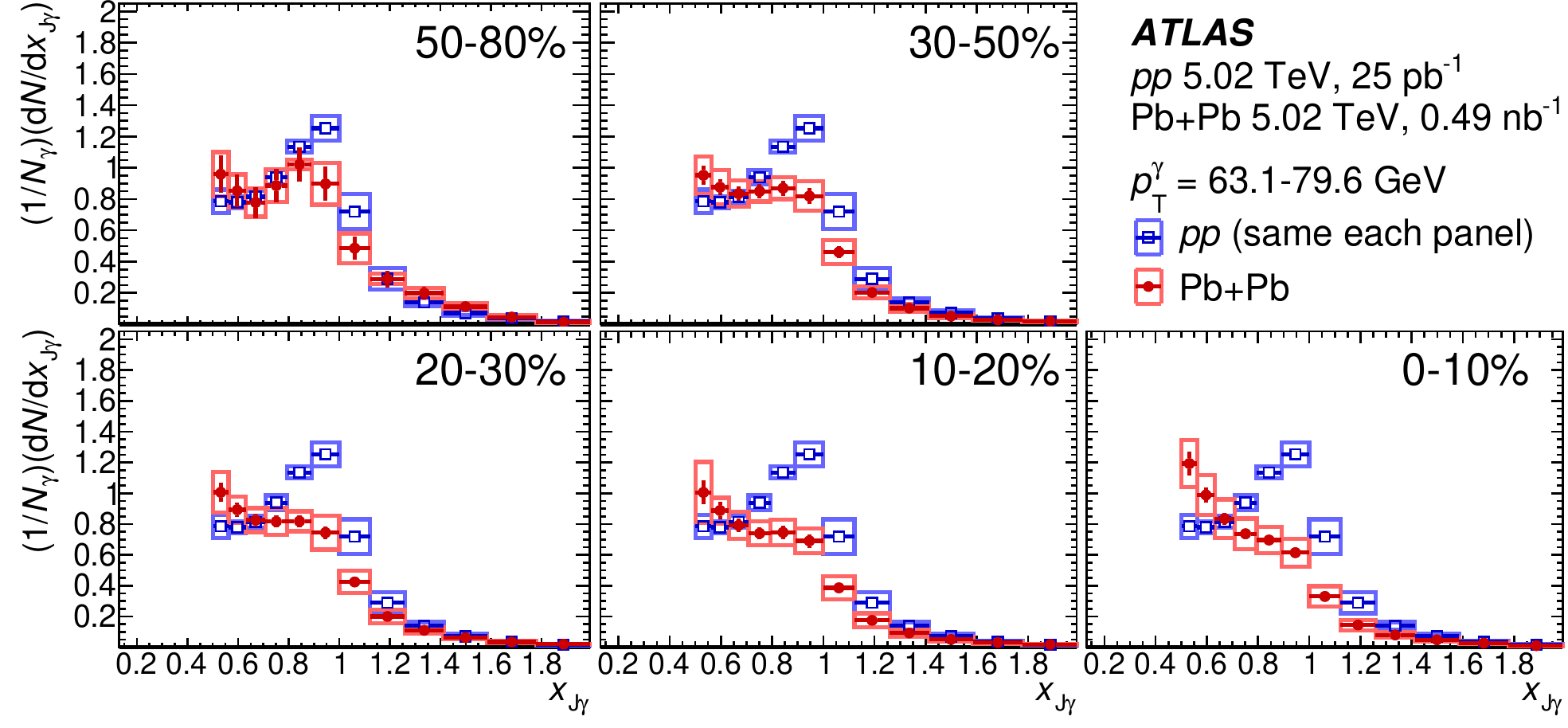}
\caption{Photon--jet \pT-balance distributions
  $(1/N_\gamma)(\mathrm{d}N/\mathrm{d}x_\mathrm{J\gamma})$ in Pb+Pb
  events~(red circles) with each panel showing a different centrality
  selection compared to that in $pp$ events (blue squares). These panels show
  results for $\pTgamma = 63.1$--$79.6$~\GeV. Total systematic
  uncertainties are shown as boxes, while statistical uncertainties
  are shown as vertical bars.
\label{fig:PbPb0}}
%\end{center}
%\end{figure*}

%\FloatBarrier

\vspace{1cm}

%\begin{figure*}[t]
%\begin{center}
\includegraphics[width=0.99\textwidth]{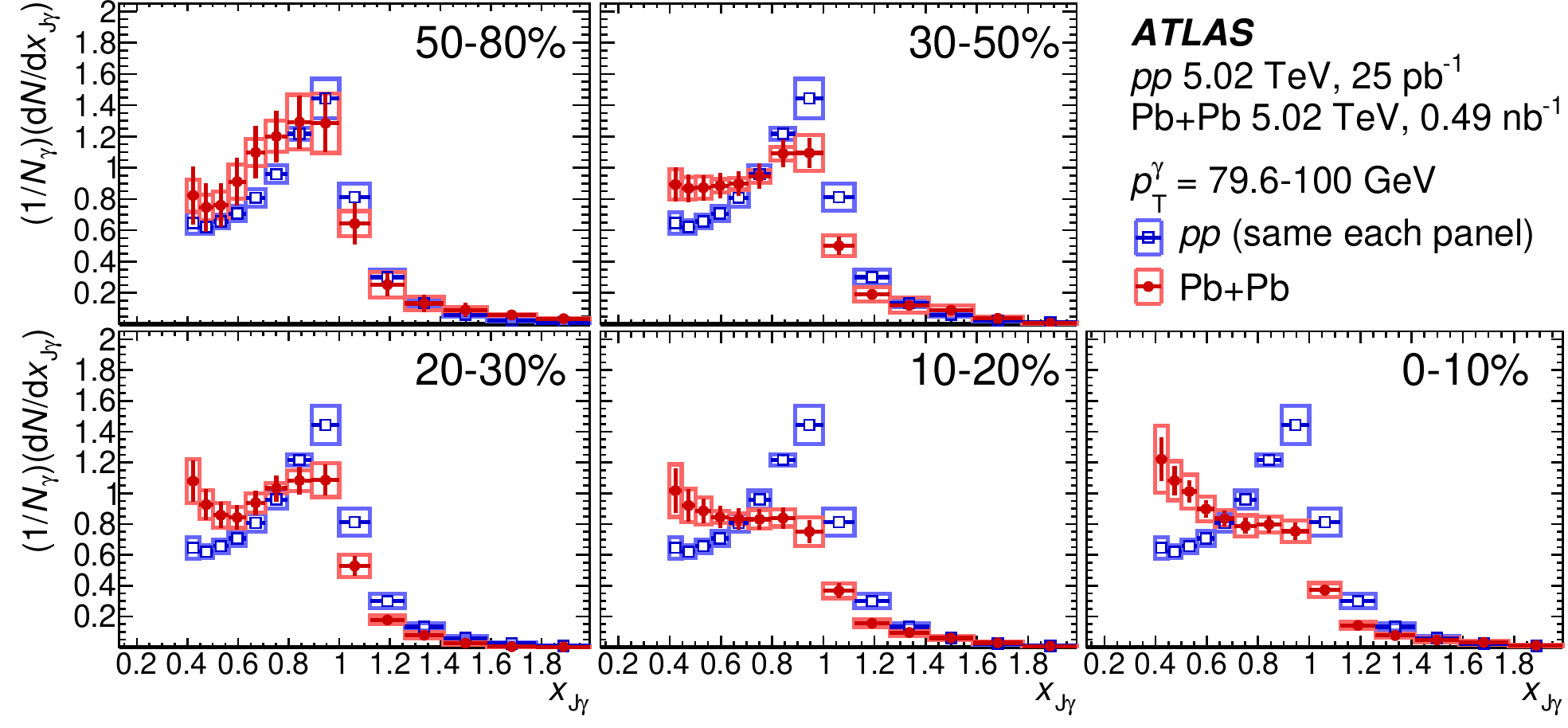}
\caption{Photon--jet \pT-balance distributions
  $(1/N_\gamma)(\mathrm{d}N/\mathrm{d}x_\mathrm{J\gamma})$ in Pb+Pb
  events~(red circles) with each panel showing a different centrality
  selection compared to that in $pp$ events (blue squares). These panels show
  results for $\pTgamma = 79.6$--$100$~\GeV. Total systematic
  uncertainties are shown as boxes, while statistical uncertainties
  are shown as vertical bars.
\label{fig:PbPb1}}
\end{center}
\end{figure*}

\begin{figure*}[p]
\begin{center}
\includegraphics[width=0.99\textwidth]{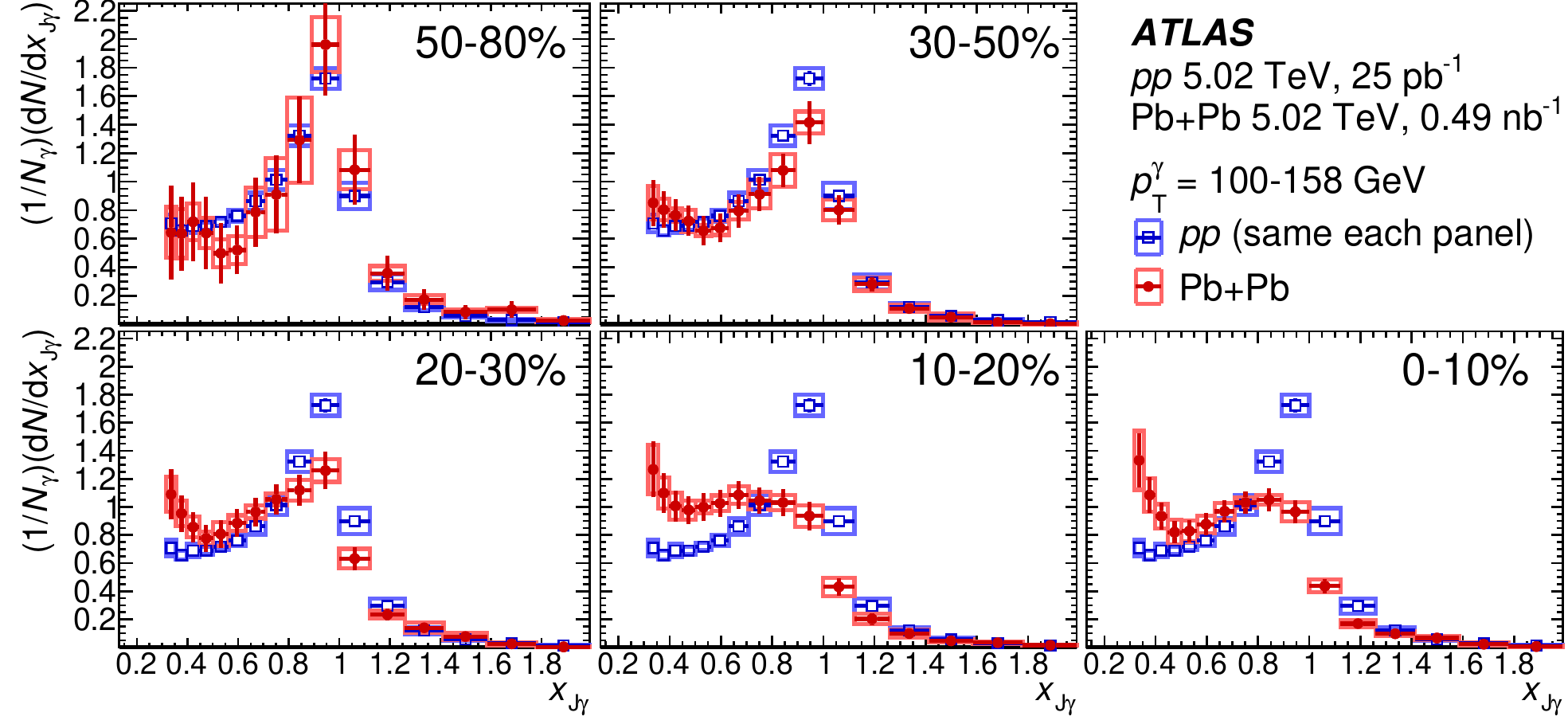}
\caption{Photon--jet \pT-balance distributions
  $(1/N_\gamma)(\mathrm{d}N/\mathrm{d}x_\mathrm{J\gamma})$ in Pb+Pb
  events~(red circles) with each panel showing a different centrality
  selection compared to that in $pp$ events (blue squares). These panels show
  results for $\pTgamma = 100$--$158$~\GeV. Total systematic
  uncertainties are shown as boxes, while statistical uncertainties
  are shown as vertical bars.
\label{fig:PbPb2}}

\vspace{1cm}
%\end{center}
%\end{figure*}

%\begin{figure*}[h]
%\begin{center}
\includegraphics[width=0.99\textwidth]{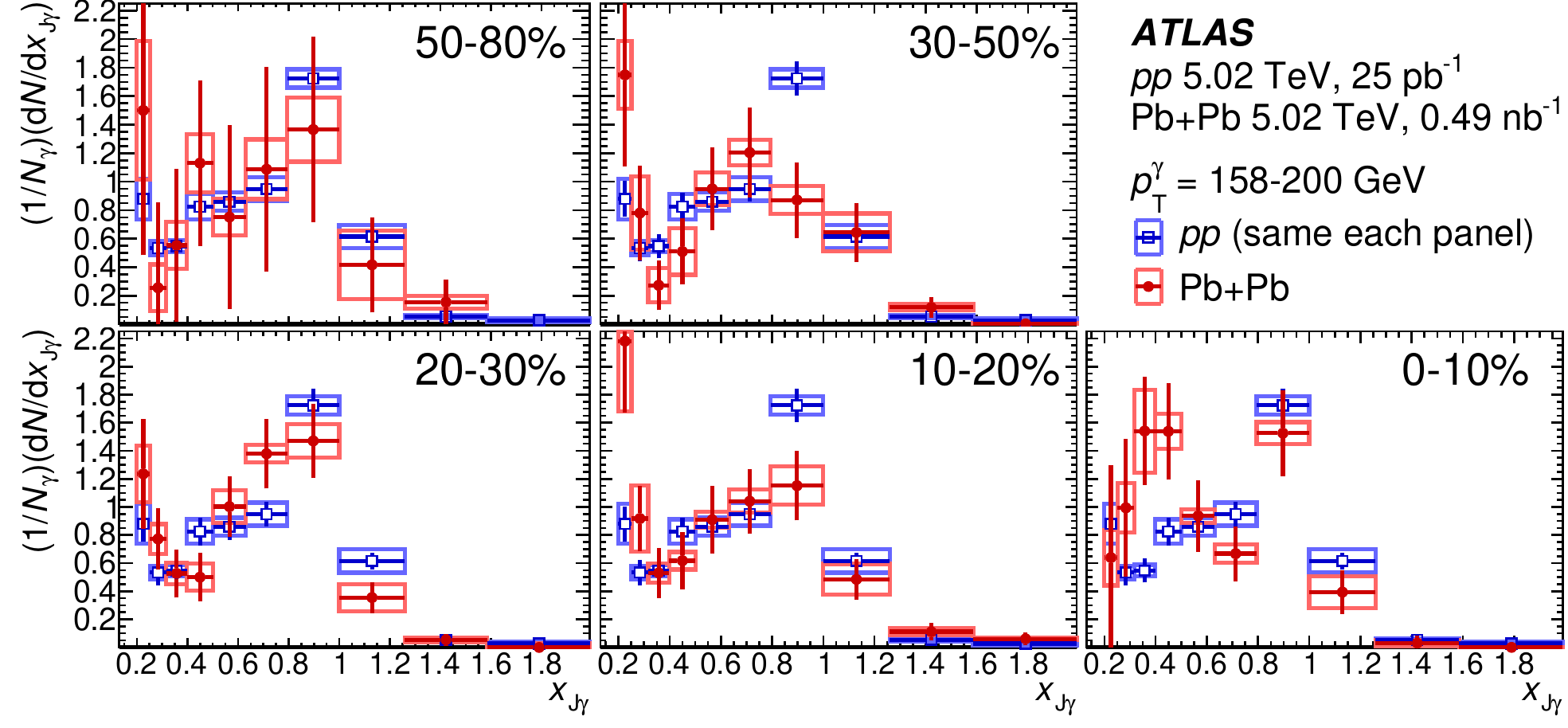}
\caption{Photon--jet \pT-balance distributions
  $(1/N_\gamma)(\mathrm{d}N/\mathrm{d}x_\mathrm{J\gamma})$ in Pb+Pb
  events~(red circles) with each panel showing a different centrality
  selection compared to that in $pp$ events (blue squares). These panels show
  results for $\pTgamma = 158$--$200$~\GeV. Total systematic
  uncertainties are shown as boxes, while statistical uncertainties
  are shown as vertical bars.
\label{fig:PbPb3}}
\vspace{1cm}
\end{center}
\end{figure*}

The unfolded $(1/N_\gamma)(\mathrm{d}N/\mathrm{d}x_\mathrm{J\gamma})$
distributions in Pb+Pb collisions are presented in
Figures~\ref{fig:PbPb0} through~\ref{fig:PbPb3}, with each figure
representing a different \pTgamma interval. Since the results are
fully corrected, they may be directly compared with the analogous \xjg
distributions in $pp$ collisions, which are reproduced in each panel
for convenience.

For all \pTgamma intervals, the \xjg distributions in Pb+Pb collisions
evolve smoothly with centrality. For peripheral collisions with
centrality 50--80\%, they are similar to those measured in $pp$
collisions. However, in increasingly more central collisions, the
distributions become progressively more modified. For the $\pTgamma <
100$~\GeV intervals shown in Figures~\ref{fig:PbPb0}
and~\ref{fig:PbPb1}, the \xjg distributions in the most central
0--10\% events are so strongly modified that they decrease
monotonically over the measured \xjg range and no peak is
observed. For the $\pTgamma > 100$~\GeV region shown in
Figure~\ref{fig:PbPb2}, the \xjg distributions retain a peak at or
near $\xjg \approx 0.9$ even in the most central collisions. However,
the magnitude of the peak is lower and significantly wider than the
sharp peak in $pp$ events. In both cases, the jet yield at small \xjg
is systematically higher than that in $pp$ collisions, by up to a
factor of two. In less central events, a peak-like structure develops
at the same position as the maximum in $pp$ events, near $\xjg \approx
0.9$. For the lowest-\pTgamma interval, this occurs only for 50--80\%
centrality events, while in the highest two \pTgamma intervals the
distribution in 0--10\% events is consistent with a local peak.
%\clearpage

%\vspace{1cm}

%\FloatBarrier

%% systematic extraction

\begin{figure*}[th]
\begin{center}
\includegraphics[width=0.49\textwidth]{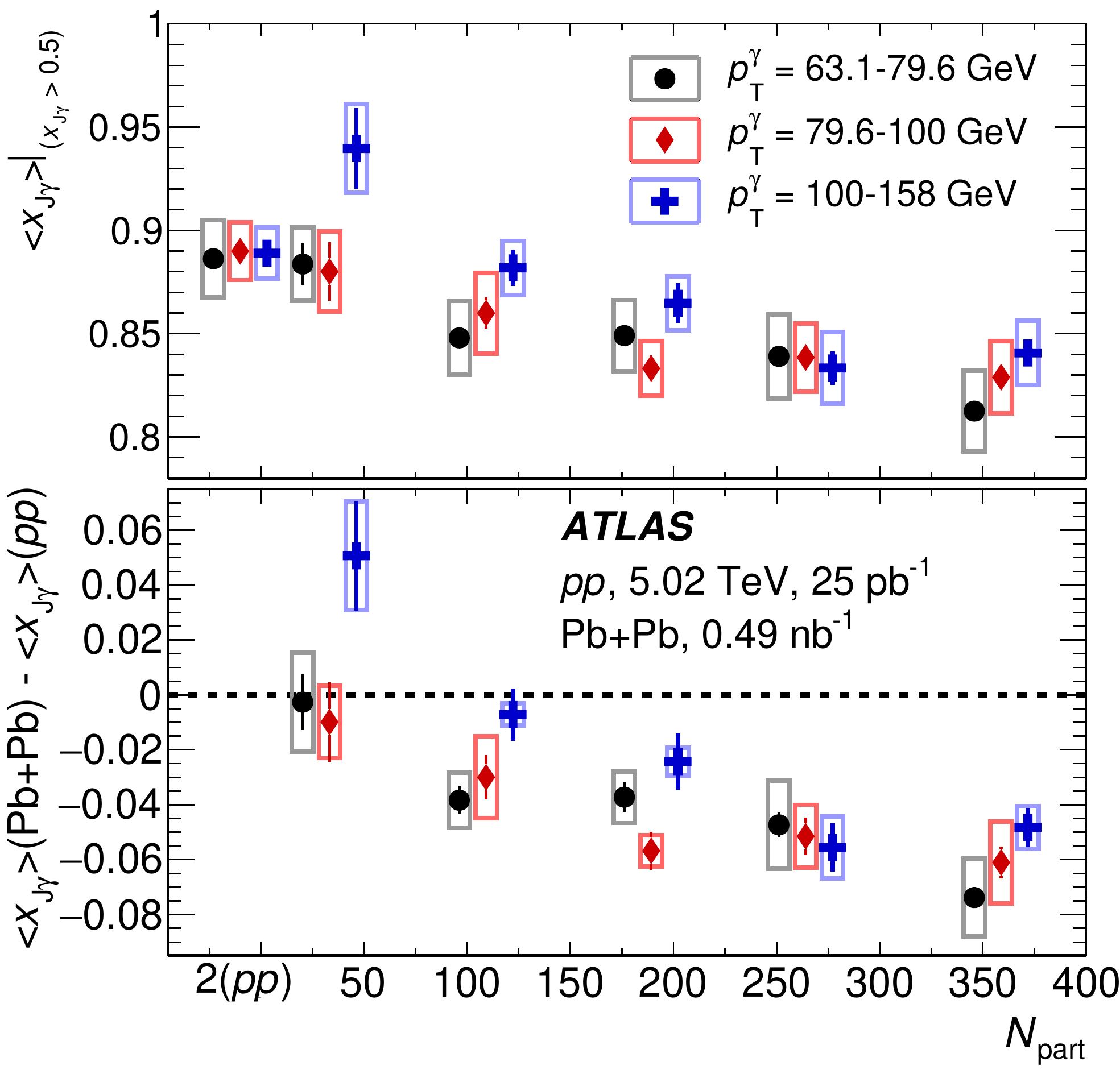}
\includegraphics[width=0.49\textwidth]{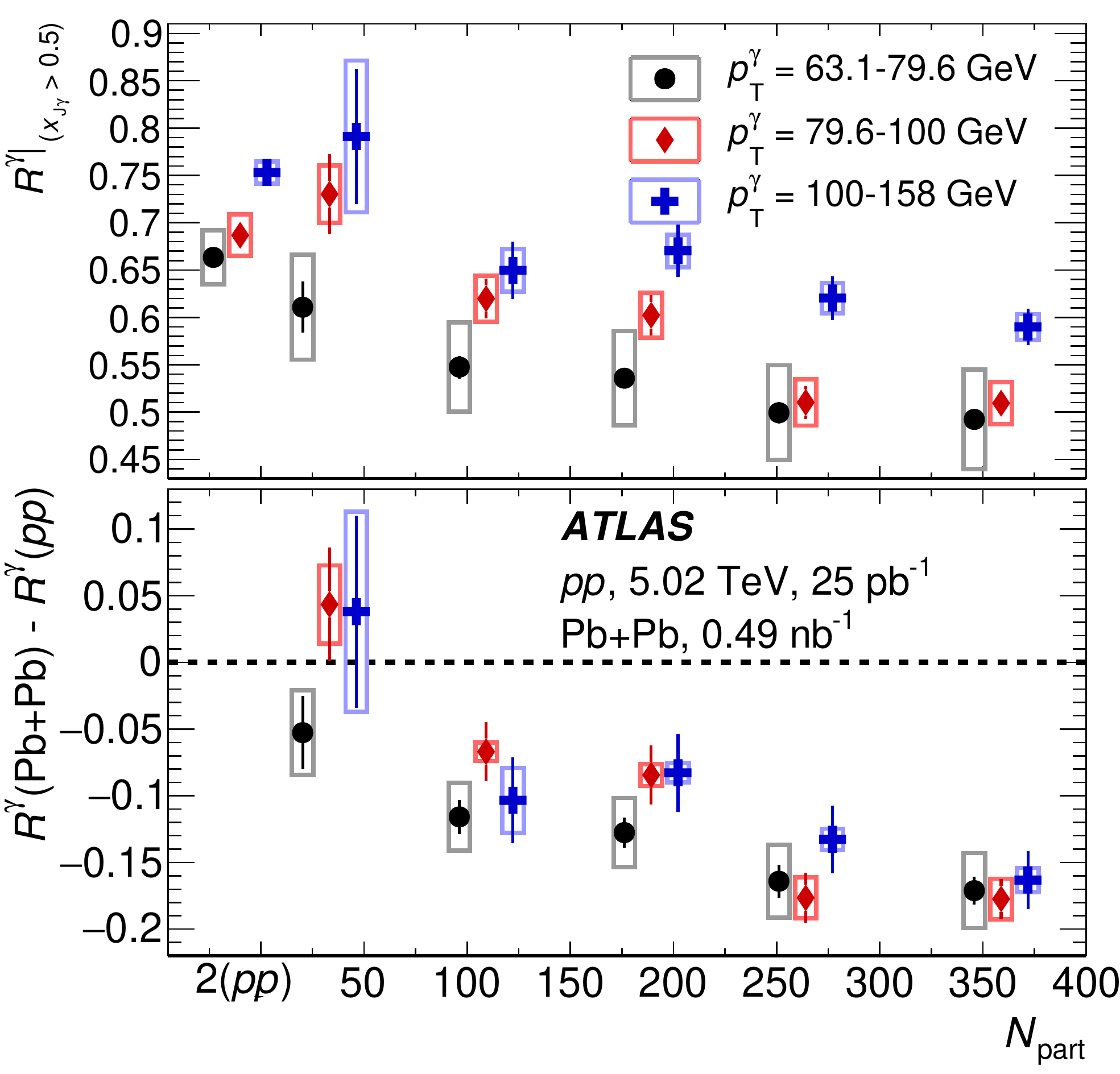}
\caption{Summary of (left) the mean jet-to-photon \pT ratio
  $\left<\xjg\right>$ and (right) the total per-photon jet yield
  $R^{\gamma}$, calculated in the region $\xjg > 0.5$. The values are
  presented as a function of the mean number of participating nucleons
  $N_\mathrm{part}$ in top panels. Each colour and symbol represents a
  different \pTgamma interval, where the lowest and highest intervals
  are displaced horizontally for clarity. The points plotted at
  $N_\mathrm{part} = 2$ correspond to $pp$ collisions. The bottom
  panels show the difference between the Pb+Pb centrality selection
  and $pp$ collisions. Boxes show the total systematic uncertainty
  while the vertical bars represent statistical uncertainties.
\label{fig:PbPb:summary}}
\end{center}
\end{figure*}

As another way of characterising how the modified \xjg distributions
depend on centrality and \pTgamma, Figure~\ref{fig:PbPb:summary}
presents their mean value, $\left<x_\mathrm{J\gamma}\right>$, and
integral, $R^{\gamma}$, with both values calculated in the region
$\xjg > 0.5$. These quantities are shown as a function of the mean
number of participating nucleons $N_{\mathrm{part}}$ in the
corresponding centrality selection, and are plotted for the first
three $\pTgamma$ intervals where they have small statistical
uncertainties. When measured in the region $\xjg > 0.5$, the value of
$\left<x_\mathrm{J\gamma}\right>$ in $pp$ collisions is observed to be
$\approx 0.89$ for all \pTgamma intervals. Simulation studies show
that, at generator level, the jet yield at $\xjg > 0.5$ corresponds to
only the leading (highest-\pT) photon-correlated jet in each
event. Thus, $\left<x_\mathrm{J\gamma}\right>$ can be interpreted as a
conditional per-jet fractional energy loss, and $R^{\gamma}$ can be
interpreted as the fraction of photons with a leading jet above $\xjg
= 0.5$. In $pp$ collisions, $R^\gamma$ ranges from $0.65$ to $0.75$ in
the three \pTgamma intervals shown, which is below unity due to the
jet selection criteria ($\Delta\phi > 7\pi/8$, $\left|\eta\right| < 2.8$).

In Pb+Pb events, $\left<x_\mathrm{J\gamma}\right>$ decreases
monotonically from the value in $pp$ collisions as the collisions
become more central. In the most central collisions, it is below the
$pp$ value by 0.04--0.06, depending on the \pTgamma interval, while in
peripheral collisions it reaches a value which is statistically
compatible with that in $pp$ events. The $R^{\gamma}$~value also
decreases monotonically as the collisions become more central,
reflecting the overall shift of the \xjg value of leading jets below
$\xjg = 0.5$. At low \pTgamma in central Pb+Pb collisions,
$R^{\gamma}$ reaches the value of $0.5$, which is only $\approx75$\%
of its value in $pp$ collisions.

%% theory comparisons

\begin{figure*}[t]
\begin{center}
\includegraphics[width=0.49\textwidth]{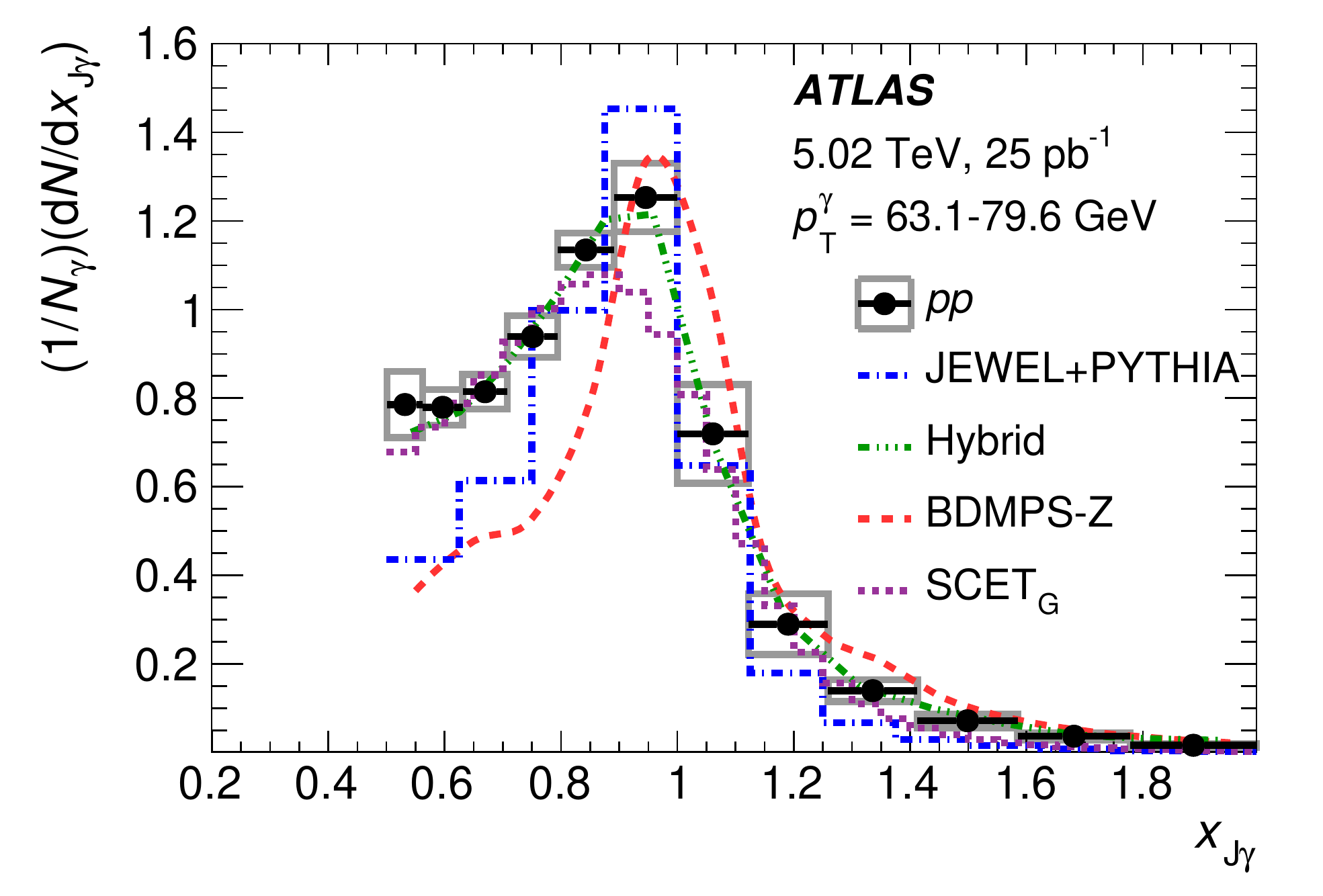}
\includegraphics[width=0.49\textwidth]{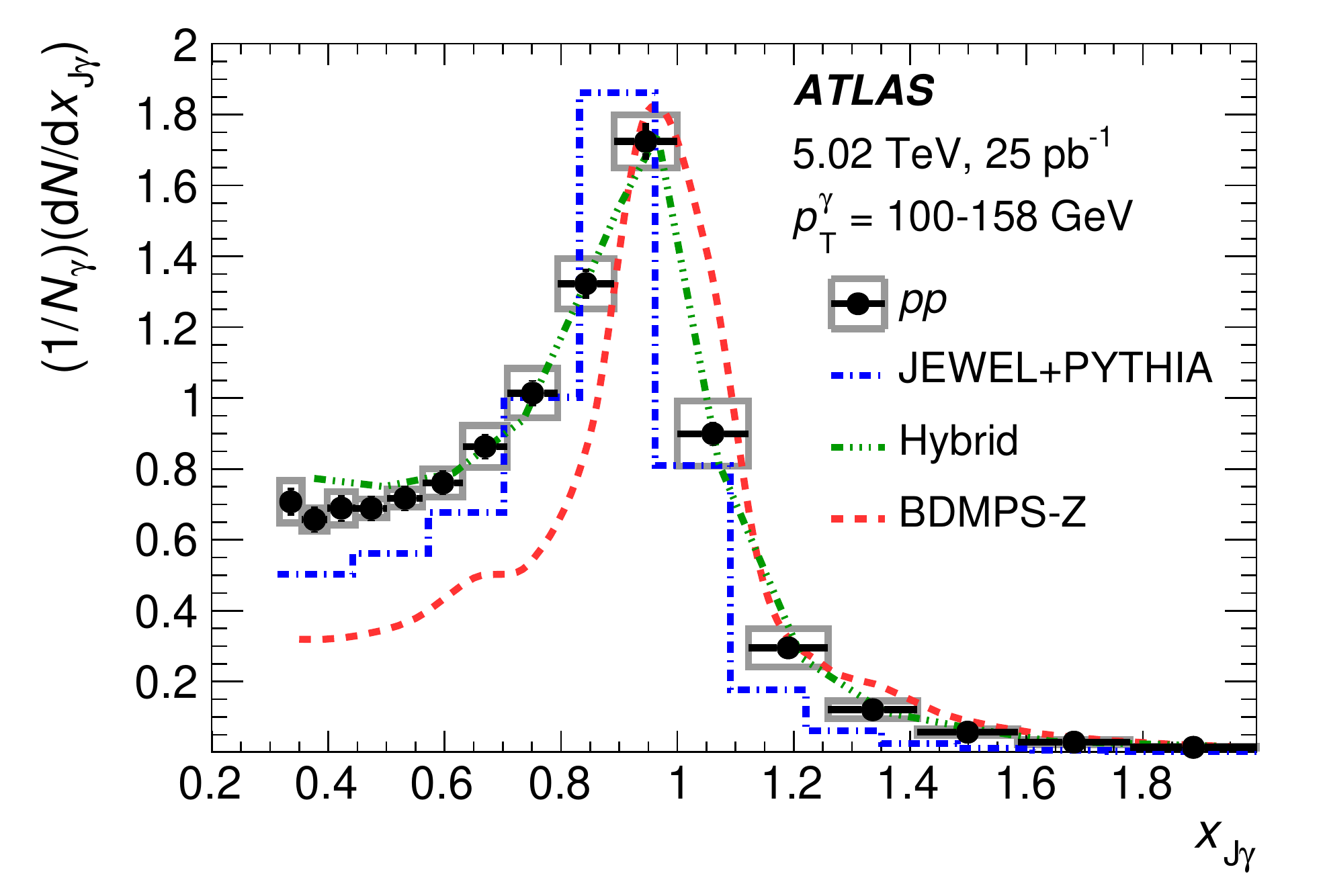}
\caption{Photon--jet \pT-balance distributions
  $(1/N_\gamma)(\mathrm{d}N/\mathrm{d}x_\mathrm{J\gamma})$ in $pp$
  collisions for (left) $\pTgamma = 63.1$--$79.6$~\GeV\ and (right)
  $\pTgamma = 100$--$158$~\GeV. The unfolded results are compared with
  the theoretical calculations shown as dashed coloured lines (see
  text). Total systematic uncertainties are shown as boxes, while
  statistical uncertainties are shown as vertical bars.
\label{fig:theory:0}}
\end{center}
\end{figure*}

The results are compared with the following theoretical predictions
which include Monte Carlo generators and analytical calculations of
jet energy loss: (1) a pQCD calculation which includes Sudakov
resummation to describe the vacuum distributions and energy loss in
Pb+Pb collisions as described in the BDMPS-Z
formalism~\cite{Chen:2018fqu}, (2) a perturbative calculation within
the framework of soft-collinear effective field theory with Glauber
gluons (SCET$_\mathrm{G}$) in the soft gluon emission (energy-loss)
limit~\cite{Kang:2017xnc}, (3) the \textsc{JEWEL} Monte Carlo event
generator which simulates QCD jet evolution in heavy-ion collisions
and includes energy-loss effects from radiative and elastic scattering
processes~\cite{KunnawalkamElayavalli:2016ttl}, and (4) the Hybrid
Strong/Weak Coupling model~\cite{Casalderrey-Solana:2015vaa} which
combines initial production using \PYTHIA with a parameterisation of
energy loss derived from holographic methods, and includes
back-reaction effects.

\begin{figure*}[t]
\begin{center}
\includegraphics[width=0.49\textwidth]{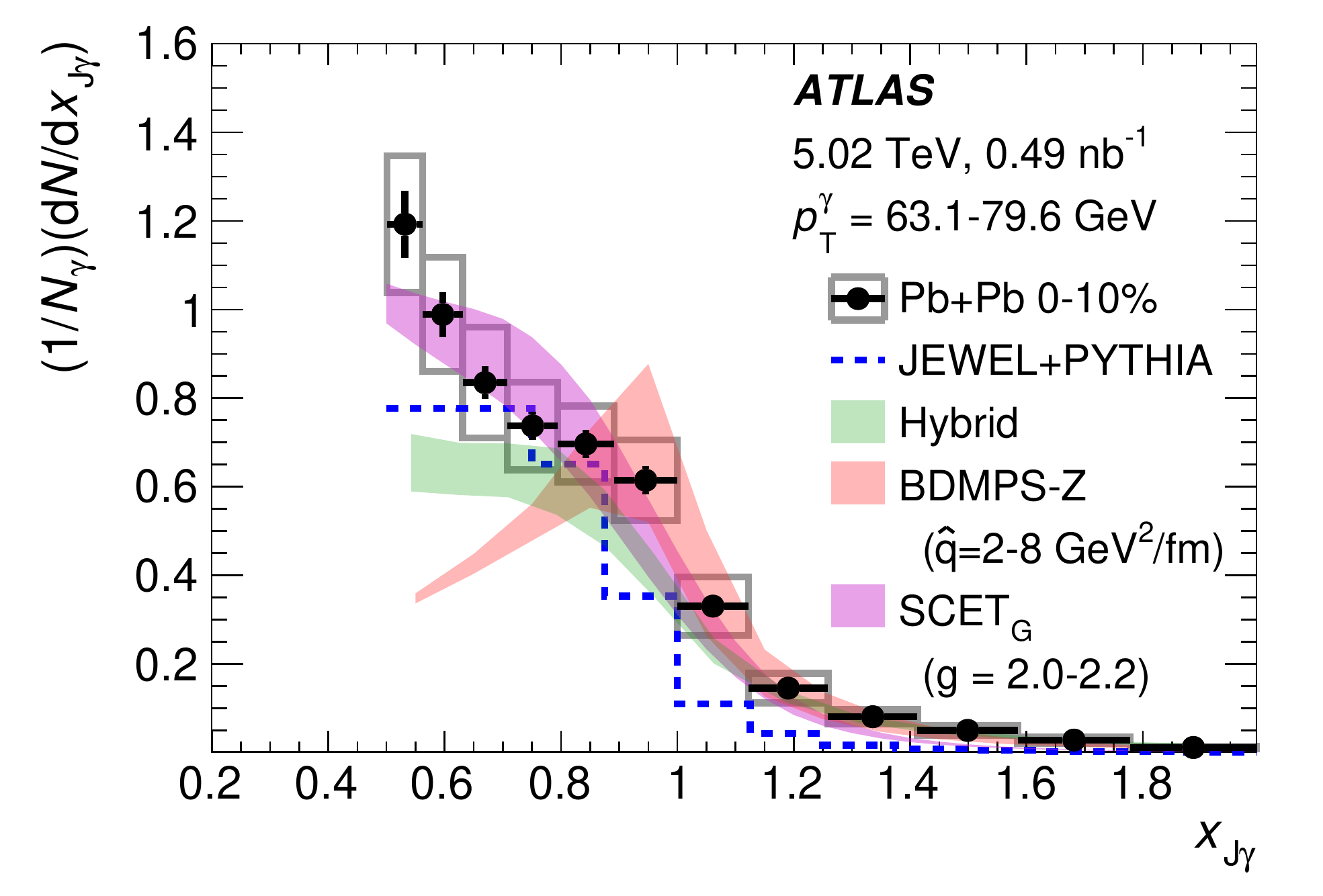}
\includegraphics[width=0.49\textwidth]{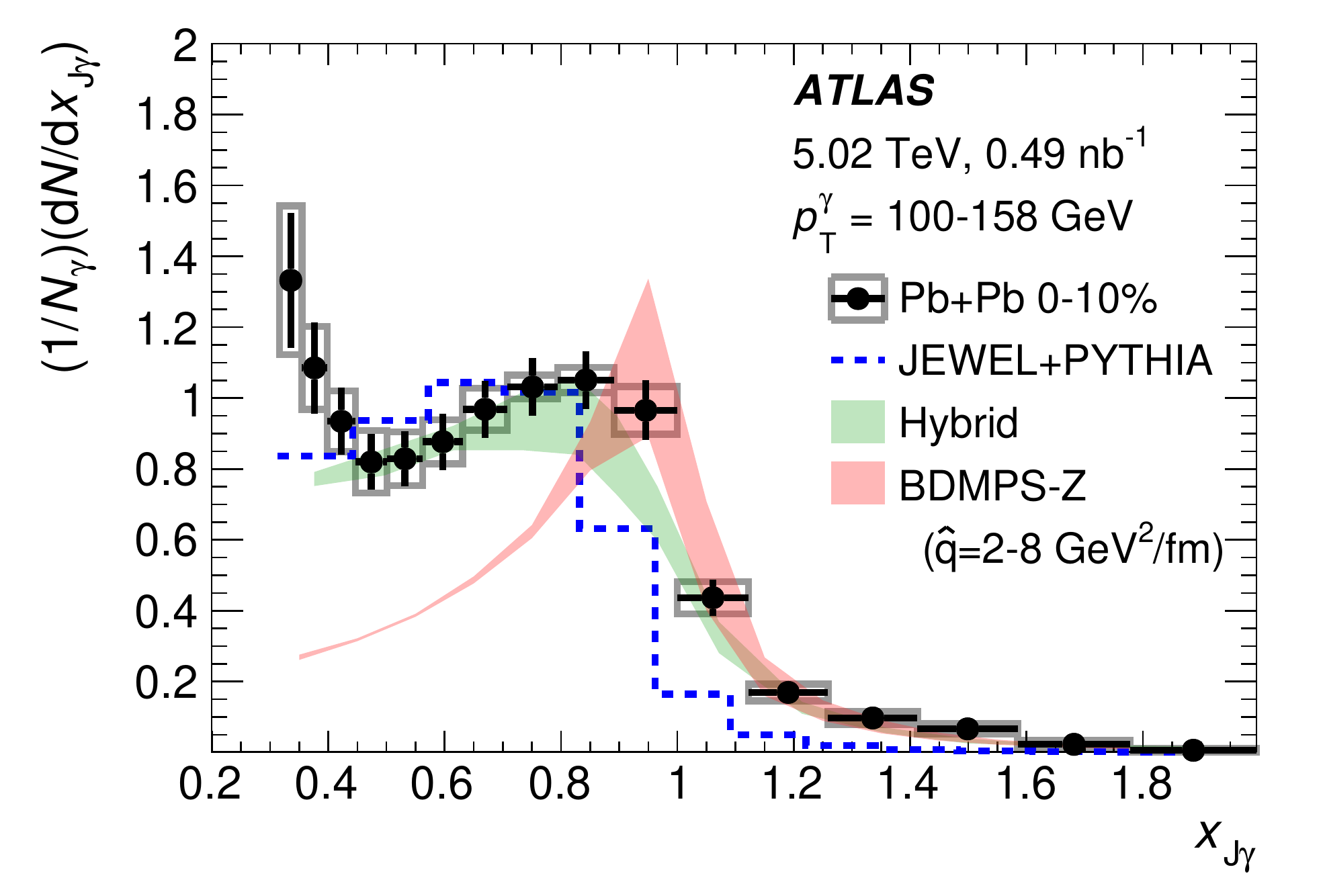}
\caption{Photon--jet \pT-balance distributions
  $(1/N_\gamma)(\mathrm{d}N/\mathrm{d}x_\mathrm{J\gamma})$ in 0--10\%
  Pb+Pb collisions for (left) $\pTgamma = 63.1$--$79.6$~\GeV\ and
  (right) $\pTgamma = 100$--$158$~\GeV. The unfolded results are
  compared with the theoretical calculations shown as dashed coloured
  lines denoting central values or coloured bands which correspond to a
  range of theoretical parameters (see text). Total systematic
  uncertainties are shown as boxes, while statistical uncertainties
  are shown as vertical bars.
\label{fig:theory:1}}
\end{center}
\end{figure*}

Figures~\ref{fig:theory:0} and~\ref{fig:theory:1} compare a selection
of the measured \xjg distributions with the results of these
theoretical predictions, where possible. Before testing the
description of energy-loss effects in Pb+Pb events, the predicted \xjg
distributions are compared with $pp$ data in
Figure~\ref{fig:theory:0}. The Hybrid model and \textsc{JEWEL}, which
use \PYTHIA for the photon--jet production in vacuum, give a good
description of $pp$ events over the measured \xjg range in both
\pTgamma intervals shown. The BDMPS-Z and SCET$_{\mathrm{G}}$
perturbative calculations capture the general features but predict
distributions that are more and less peaked, respectively, than those
in data.

In Pb+Pb events with low \pTgamma, shown in the left panel of
Figure~\ref{fig:theory:1}, the \text{JEWEL}, Hybrid, and
SCET$_{\mathrm{G}}$ models successfully capture several key features
of the \xjg distribution, including the absence of a visible peak, and
the monotonically increasing behaviour with decreasing \xjg. The
BDMPS-Z model predicts a suppression of the yield near $\xjg \approx
0.9$ relative to what is predicted in $pp$ events, consistent with the
trend in data. However, it underestimates the yield at low \xjg in
both $pp$ and Pb+Pb collisions. In the higher-\pTgamma interval, the
Hybrid model and \text{JEWEL} successfully describe the reappearance
of a localised peak near $\xjg \approx 0.9$. However, none of the
models considered here describe the increase of the jet yield at $\xjg
< 0.5$ above that observed in $pp$ events. Additional comparisons
between these data and theoretical calculations which are differential
in both \pTgamma and centrality will further constrain the description
of the strongly coupled medium in these models.

\section{Conclusion}

This Letter presents a study of photon--jet transverse momentum
correlations for photons with $63.1 < \pTgamma <
200$~\GeV in Pb+Pb collisions at $\sqrt{s_\mathrm{NN}} =
5.02$~\TeV\ and $pp$ collisions at $\sqrt{s}=5.02$~\TeV. The data were
recorded with the ATLAS detector at the LHC and correspond to
integrated luminosities of $0.49$~nb$^{-1}$ and $25$~pb$^{-1}$,
respectively. The data are corrected for the presence of combinatoric
photon--jet pairs and of dijet pairs where one of the jets is
misidentified as a photon. The measured quantities in data are fully
corrected for detector effects and reported at the particle
level. Per-photon distributions of the jet-to-photon \pT ratio, $\xjg
= \pTjet / \pTgamma$, are measured for
pairs with an azimuthally balanced configuration, $\Delta\phi >
7\pi/8$. In $pp$ events, the data are well reproduced by event
generators or models that depend on them, but are not fully described
in detail by approaches based on perturbative calculations.

In Pb+Pb collisions, \xjg distributions are observed to have a
significantly modified total yield and shape compared with those in
$pp$ collisions. These modifications have a smooth onset as a function
of Pb+Pb event centrality and \pTgamma. In peripheral collisions at
high \pTgamma, the distributions in Pb+Pb are statistically compatible
with those in $pp$. In the most central Pb+Pb events at low \pTgamma,
the yield decreases monotonically with increasing \xjg over the
measured range, in strong contrast to the sharply peaked distributions
in $pp$ events. However, in less central events or in higher-\pTgamma
intervals, the \xjg distributions retain a peak-like excess at an \xjg
value similar to that in $pp$ collisions but with a smaller per-photon
yield. This last observation suggests that the amount of energy lost
by jets in single events has a broad distribution, with a small but
significant population of jets retaining a $pp$-like \pT correlation
with the photon because they do not lose an appreciable amount of
energy.

These results are sensitive to how partons initially produced opposite
to a high-\pT photon lose energy in their interactions with the hot
nuclear medium. Taken together with other measurements of single-jet
and dijet production, the data provide new, complementary information
about how energy loss in the strongly coupled medium varies with the
initial parton flavour and \pT.

%-------------------------------------------------------------------------------
\section*{Acknowledgements}
%-------------------------------------------------------------------------------

% Acknowledgements for papers with collision data
% Version 24-Oct-2018

% Standard acknowledgements start here
%----------------------------------------------
We thank CERN for the very successful operation of the LHC, as well as the
support staff from our institutions without whom ATLAS could not be
operated efficiently.

We acknowledge the support of ANPCyT, Argentina; YerPhI, Armenia; ARC, Australia; BMWFW and FWF, Austria; ANAS, Azerbaijan; SSTC, Belarus; CNPq and FAPESP, Brazil; NSERC, NRC and CFI, Canada; CERN; CONICYT, Chile; CAS, MOST and NSFC, China; COLCIENCIAS, Colombia; MSMT CR, MPO CR and VSC CR, Czech Republic; DNRF and DNSRC, Denmark; IN2P3-CNRS, CEA-DRF/IRFU, France; SRNSFG, Georgia; BMBF, HGF, and MPG, Germany; GSRT, Greece; RGC, Hong Kong SAR, China; ISF and Benoziyo Center, Israel; INFN, Italy; MEXT and JSPS, Japan; CNRST, Morocco; NWO, Netherlands; RCN, Norway; MNiSW and NCN, Poland; FCT, Portugal; MNE/IFA, Romania; MES of Russia and NRC KI, Russian Federation; JINR; MESTD, Serbia; MSSR, Slovakia; ARRS and MIZ\v{S}, Slovenia; DST/NRF, South Africa; MINECO, Spain; SRC and Wallenberg Foundation, Sweden; SERI, SNSF and Cantons of Bern and Geneva, Switzerland; MOST, Taiwan; TAEK, Turkey; STFC, United Kingdom; DOE and NSF, United States of America. In addition, individual groups and members have received support from BCKDF, CANARIE, CRC and Compute Canada, Canada; COST, ERC, ERDF, Horizon 2020, and Marie Sk{\l}odowska-Curie Actions, European Union; Investissements d' Avenir Labex and Idex, ANR, France; DFG and AvH Foundation, Germany; Herakleitos, Thales and Aristeia programmes co-financed by EU-ESF and the Greek NSRF, Greece; BSF-NSF and GIF, Israel; CERCA Programme Generalitat de Catalunya, Spain; The Royal Society and Leverhulme Trust, United Kingdom. 

The crucial computing support from all WLCG partners is acknowledged gratefully, in particular from CERN, the ATLAS Tier-1 facilities at TRIUMF (Canada), NDGF (Denmark, Norway, Sweden), CC-IN2P3 (France), KIT/GridKA (Germany), INFN-CNAF (Italy), NL-T1 (Netherlands), PIC (Spain), ASGC (Taiwan), RAL (UK) and BNL (USA), the Tier-2 facilities worldwide and large non-WLCG resource providers. Major contributors of computing resources are listed in Ref.~\cite{ATL-GEN-PUB-2016-002}.
%----------------------------------------------

%----------------------------------------------

%-------------------------------------------------------------------------------
\clearpage

%-------------------------------------------------------------------------------

%-------------------------------------------------------------------------------
% If you use biblatex and either biber or bibtex to process the bibliography
% just say \printbibliography here

% dvp: for the paper: 
\printbibliography

% If you want to use the traditional BibTeX you need to use the syntax below.
%\bibliographystyle{bib/bst/atlasBibStyleWoTitle}
%\bibliography{atlas-document,bib/ATLAS,bib/CMS,bib/ConfNotes,bib/PubNotes}
%-------------------------------------------------------------------------------

\clearpage 

% ATLAS Collaboration author list
% Reference date of HION-2018-06 is 2018-04-27
% Author list last updated on date 13-JUL-18
% Data extracted on 13-Jul-2018 for paper reference HION-2018-06
% at 2:18pm
 
\begin{flushleft}
{\Large The ATLAS Collaboration}

\bigskip

M.~Aaboud$^\textrm{\scriptsize 34d}$,    
G.~Aad$^\textrm{\scriptsize 99}$,    
B.~Abbott$^\textrm{\scriptsize 124}$,    
O.~Abdinov$^\textrm{\scriptsize 13,*}$,    
B.~Abeloos$^\textrm{\scriptsize 128}$,    
D.K.~Abhayasinghe$^\textrm{\scriptsize 91}$,    
S.H.~Abidi$^\textrm{\scriptsize 164}$,    
O.S.~AbouZeid$^\textrm{\scriptsize 39}$,    
N.L.~Abraham$^\textrm{\scriptsize 153}$,    
H.~Abramowicz$^\textrm{\scriptsize 158}$
% ...
\bigskip
\\

% ...
$^{13}$Institute of Physics, Azerbaijan Academy of Sciences, Baku; Azerbaijan.\\
% ...
$^{34}$$^{(a)}$Facult\'e des Sciences Ain Chock, R\'eseau Universitaire de Physique des Hautes Energies - Universit\'e Hassan II, Casablanca;$^{(b)}$Centre National de l'Energie des Sciences Techniques Nucleaires (CNESTEN), Rabat;$^{(c)}$Facult\'e des Sciences Semlalia, Universit\'e Cadi Ayyad, LPHEA-Marrakech;$^{(d)}$Facult\'e des Sciences, Universit\'e Mohamed Premier and LPTPM, Oujda;$^{(e)}$Facult\'e des sciences, Universit\'e Mohammed V, Rabat; Morocco.\\
% ...
$^{39}$Niels Bohr Institute, University of Copenhagen, Copenhagen; Denmark.\\
% ...
$^{91}$Department of Physics, Royal Holloway University of London, Egham; United Kingdom.\\
% ...
$^{99}$CPPM, Aix-Marseille Universit\'e, CNRS/IN2P3, Marseille; France.\\
% ...
$^{124}$Homer L. Dodge Department of Physics and Astronomy, University of Oklahoma, Norman OK; United States of America.\\
% ...
$^{128}$LAL, Universit\'e Paris-Sud, CNRS/IN2P3, Universit\'e Paris-Saclay, Orsay; France.\\
% ...
$^{153}$Department of Physics and Astronomy, University of Sussex, Brighton; United Kingdom.\\
% ...
$^{158}$Raymond and Beverly Sackler School of Physics and Astronomy, Tel Aviv University, Tel Aviv; Israel.\\
% ...
$^{164}$Department of Physics, University of Toronto, Toronto ON; Canada.\\
% ...
$^{a}$ Also at Borough of Manhattan Community College, City University of New York, NY; United States of America.\\
$^{b}$ Also at Centre for High Performance Computing, CSIR Campus, Rosebank, Cape Town; South Africa.\\
$^{c}$ Also at CERN, Geneva; Switzerland.\\
$^{d}$ Also at CPPM, Aix-Marseille Universit\'e, CNRS/IN2P3, Marseille; France.\\
$^{e}$ Also at D\'epartement de Physique Nucl\'eaire et Corpusculaire, Universit\'e de Gen\`eve, Gen\`eve; Switzerland.\\
% ...
$^{*}$ Deceased

\end{flushleft}

% Created with Glance <Atlas.Glance@cern.ch>

\end{document}